\documentclass[11pt]{article}
\usepackage{graphicx}
\usepackage{amssymb}
\usepackage{amsmath}
\usepackage{algpseudocode}
\usepackage{algorithm}

\newcommand{\bp}{\bar{p}_{\tau}}
\newcommand{\bx}{\bar{\xi}_{\tau}}
\newcommand{\mP}{\mathcal{P}_{\tau}}

\title{A discrete spacetime model for quantum mechanics}
\author{Antonio Sciarretta}
\date{}
\begin{document}
\maketitle
\section*{Abstract}

This paper presents a simple model that mimics quantum mechanics (QM) results in terms of probability fields of free particles subject to self-interference, without using Schr\"{o}dinger equation or complex wavefunctions. Unlike the standard QM picture, the proposed model only uses integer-valued quantities and arithmetic operations. In particular, it assumes a discrete spacetime under the form of an euclidean lattice. The proposed approach describes individual particle trajectories as random walks. Transition probabilities are simple functions of a few quantities that are either randomly associated to the particles during their preparation, or stored in the lattice sites they visit during the walk. Non-relativistic QM predictions, particularly self-interference, are retrieved as probability distributions of similarly-prepared ensembles of particles. Extension to interacting particles is discussed but not detailed in this paper.

%%%%%%%%%%%%%%%%%%%%%%%%%%%%%%%%%%%%%%%%%%%%%%%%%
%%%%%%%%%%%%%%%%%%%%%%%%%%%%%%%%%%%%%%%%%%%%%%%%%
\section{Introduction}
%%%%%%%%%%%%%%%%%%%%%%%%%%%%%%%%%%%%%%%%%%%%%%%%%
%%%%%%%%%%%%%%%%%%%%%%%%%%%%%%%%%%%%%%%%%%%%%%%%%

Despite being an extremely succesful theory to predict the outcome of experiments with particles and other microscopic objects, quantum mechanics (QM) is believed by many to be incomplete or, at least, not fully understood. In particular, the ``strange" or non-classical phenomena of QM, like self-interference and Born's rule, are described in terms of abstract mathematical objects. Although some have been ready to interpret the complex-valued wavefunction as a real object, wavefunctions are generally seen as mathematical tools serving to calculate probabilities from their square moduli. Contrasting to real-valued mathematics and one-to-one mapping between real variables and observables of classical theories, the standard description is thus sometimes considered as purely operational.

In this paper a model is presented that, although very simple in terms of mathematical development and formalism, seems to be able to predict probability fields of at least free particles, and particularly self-interference, without using Schr\"{o}dinger equation, wavefunctions, or path integrals. The proposed model assumes a discrete spacetime under the form of a euclidean lattice. Individual particle trajectories are described as random walks. Transition probabilities are simple functions of a few quantities that are either randomly associated to the particles during their preparation, or stored in the lattice sites they visit during the walk. Non-relativistic QM predictions are retrieved as probability distributions of similarly-prepared ensembles of particles. Unlike the standard QM picture \cite{feynmann}, the proposed model only uses integer-valued quantities and arithmetic operations, avoiding complex-valued operations on the random walks.

The proposed model is different from ensemble interpretations \cite{ballentine, neumaier} in the sense that it also describes the behavior of individual particles. With respect to De Broglie--Bohm mechanics \cite{bohm} and deterministic trajectory representation \cite{floyd}, the proposed model introduces a non-deterministic behavior but does not appeal to non-localities.

Intrinsic indeterminism is already contained in stochastic interpretations of QM that are, however, mostly aimed at retrieving the Schr\"{o}dinger equation from a classical equation of motion plus a stochastic force. The Born probability rule remains unexplained in this context \cite{nelson, fritsche, carroll}, or is founded on the definition of probability density of particles as the squared intensity of an associated wave \cite{groessing}. This latter assumption is not used in the proposed approach, which in contrast predicts nonclassical consequences of Born rule (as double-slit interference) only from the random walk features. However, unlike other corpuscolar models \cite{michielsen} the proposed model is not restricted to double-slit intereference as it is not explicitly based on ad-hoc trigonometric functions inspired by the complex waveforms structure.
%\cite{nelson, davidson, fliess, fritsche, carroll,} 

The idea of lattice or discrete-time algorithms that reproduce particle propagation in the continuum limit is also not new. However, the proposed model uses the lattice only as the support for particle motion, not for wavefunctions or other mathematical operators as, e.g., in \cite{dowker, bialynicki, deraedt}. While other random walks or spacetime quantizations \cite{ord, janaswamy, badiali} are able to reproduce the emergence of Schr\"{o}dinger equation from pure combinatorics, again the Born probability rule and thus interference are not explained in such models, while naturally emerges in the proposed one. In order to reproduce interference, antiparticles are not appealed to, as in some abstract lattice gas models for waves \cite{chen}, nor negative probabilities.

The paper is organized as follows. Section~2 presents the assumptions concerning spacetime, which naturally lead to Heisemberg's uncertainty principle. Section~3 describes the model for free motion without quantum forces and retrieves de Broglie relation and Schr\"{o}dinger equation. Section~4 describes the model with quantum forces and shows numerical results for several self-interference scenarios. Section~5 is just a short introduction to possible extensions to treat interacting particles and the transition to relativistic QM.

%%%%%%%%%%%%%%%%%%%%%%%%%%%%%%%%%%%%%%%%%%%%%%%%%
%%%%%%%%%%%%%%%%%%%%%%%%%%%%%%%%%%%%%%%%%%%%%%%%%
\section{The lattice: uncertainty principle}
%%%%%%%%%%%%%%%%%%%%%%%%%%%%%%%%%%%%%%%%%%%%%%%%%
%%%%%%%%%%%%%%%%%%%%%%%%%%%%%%%%%%%%%%%%%%%%%%%%%

The proposed model assumes that the spacetime is inherently discrete. Limiting for simplicity the analysis to one dimension $x$, that means that only values $x=\xi X$, $\xi \in \mathbb{Z}$ and $t=\tau T$, $\tau \in \mathbb{N}$ are meaningful. Noninteger values of space and time are simply impossible in this picture. The two fundamental quantities $X$ and $T$ are the size of the lattice that constitutes the space and the fundamental temporal resolution, respectively.

Under this assumption, a particle's history consists of a succession of points $\{\xi_n,\tau_n\}$ in the spacetime, where $n\in \mathbb{Z}$ is the discrete index that describes advance in history, here denoted as ``iteration". Advance in time is unidirectional and unitary, that is, $\tau+1$ follows necessarily $\tau$. Advance in space is still unitary but bidirectional. If in a certain iteration a particle resides at the location $\xi$ of the spatial lattice, in the next iteration the particle can only reside at locations $\xi+1$, $\xi$, or $\xi-1$. The local velocity of this mechanism, $\mathcal{V}$, is a random variable that can take only the three discrete values $\upsilon=\{+1, 0, -1\}$, as described in Sect.~\ref{sec:single} and Sect.~\ref{sec:interf}.

%The instantaneous velocity in physical units is therefore $v=(\beta+\gamma) \displaystyle\frac{X}{T}$. 

%Consider for the moment only free motion, without interference, such that $\xi_{n+1}=\xi_{n}+\upsilon_{n}$. The observable velocity of the particle as the result of a observation process lasting $N$ iterations or time steps ($N$ is arbitrary) would be
%%
%\begin{equation}
  %\bar{v} := \frac{1}{N} \sum_{n=1}^N \upsilon_{n} \frac{X}{T}.
  %\label{eqn:v}
%\end{equation}
%%
The maximum velocity that a particle can reach is the speed of light $c$. Light trajectory in the positive direction corresponds to $\upsilon_{n}=1$, $\forall n$. Consequently, one constraint to the fundamental lattice quantities is necessarily
\begin{equation}
  \frac{X}{T} = c.
  \label{eqn:cc} 
\end{equation}

The average velocity could only be estimated by an observer as a sample mean of successive $\upsilon$'s. Every observation lasting $N$ iterations will give an approximation $\bar{v}$ of the ``true" average velocity. Consider, e.g., $N=1$. The observed velocity can be $+c$, $0$ or $-c$. Thus the uncertainty on $\bar{v}$ is $c$ in absolute value. For $N=2$, the possible outcomes for the sample mean are $c$, $c/2$, $0$, $-c/2$, and $-c$. Thus the uncertainty on $\bar{v}$ is $c/2$ in absolute value. Extending these considerations, the uncertainty on $\bar{v}$ after an observation lasting $N$ iterations is $c/N$ in lattice units.

Moreover, an observation lasting $N$ iterations necessarily implies a change in the position of the particle. The span of the particle during the observation ranges from $NX$ to $-NX$. Thus the uncertainty on the position of the particle at the end of the observation is $2N$ in lattice units.

Using the two results above, and denoting $\Delta v(N)$ and $\Delta x(N)$ the uncertainties of velocity and position as a function of observation horizon $N$, the relationship
\begin{equation}
  \Delta v(N) \cdot \Delta x(N) = \frac{c}{N} \cdot 2NX = \frac{2X^2}{T}
  \label{eqn:heis}
\end{equation}
holds.

The latter equation resembles the Heisenberg uncertainty principle since it fixes an inverse proportionality between the uncertainty with which the velocity of a particle can be known and the uncertainty with which its position can be known. Multiplying by the particle mass $m$, and comparing (\ref{eqn:heis}) to Heisenberg uncertainty principle, one obtains that the two fundamental lattice quantities are related to the Planck constant,
\begin{equation}
  m \frac{X^2}{T} = \frac{\hbar}{2} = \frac{h}{4\pi}.
  \label{eqn:hbar}
\end{equation}
The term $4\pi$ (the solid angle of a sphere) holds for three-dimensional spaces. In our example case of a one-dimensional space, this term reduces to 2, the measure of the unit 1-sphere.
%$\Omega_1=\frac{2\pi^{1/2}}{\Gamma(1/2)}$, where $\Gamma$ is the Gamma function, i.e., $\Omega_1=2$. 
Thus, combining (\ref{eqn:cc}) with the accordingly modified (\ref{eqn:hbar}), the values for the fundamental lattice quantities are obtained as 
\begin{equation}
%  X=\frac{h}{2mc}
  X=\frac{h}{2mc},
  \label{eqn:X}
\end{equation}
and
\begin{equation}
%  T = \frac{h}{2mc^2}
  T = \frac{h}{2mc^2}.
  \label{eqn:T}
\end{equation}
Note that the Compton wavelength is retrieved as twice the fundamental lattice size $X$.

The role of mass is not completely clear at this point. Likely, general relativity will serve to integrate it into the picture.

%%%%%%%%%%%%%%%%%%%%%%%%%%%%%%%%%%%%%%%%%%%%%%%%%
%%%%%%%%%%%%%%%%%%%%%%%%%%%%%%%%%%%%%%%%%%%%%%%%%
\section{Free motion without interference} \label{sec:single}
%%%%%%%%%%%%%%%%%%%%%%%%%%%%%%%%%%%%%%%%%%%%%%%%%

This section will first describe the equations of motion of a free particle in the proposed model. Then, the stochastic variables associated with the motion will be analyzed. Finally, the equivalence with the wavefunction picture and Schr\"{o}dinger equation will be retrieved.

%%%%%%%%%%%%%%%%%%%%%%%%%%%%%%%%%%%%%%%%%%%%%%%%%
\subsection{Particle dynamics} \label{sec:prop}
%%%%%%%%%%%%%%%%%%%%%%%%%%%%%%%%%%%%%%%%%%%%%%%%%

This section describes the propagation rules of a particle on the lattice, or its \emph{dynamics}. 

\subsubsection{Equations of motion} 
%%%%%%%%%%%%%%%%%%%%%%%%%%%%%%%%%%%%%%%%%%%%%%%%%
As stated in the previous section, time can only increase by one unit at each iteration. Time is re-initialized to zero whenever the particle interacts with the environment (external forces). This event is called \emph{preparation} in the following. A stochastic variable representing time at an iteration $n$ could be introduced as
\begin{equation}
  \mathcal{T}_n := \sum_{n'=n_0}^{n} 1.
	\label{eqn:tstoc}
\end{equation}
where $n_0$ is the most recent iteration when the particle has undergone preparation, that is, has interacted with the environment and has been actualized. However, $\mathcal{T}_n$ would have a deterministic distribution, $\Pr(\mathcal{T}_n=\tau)=\delta(\tau-(n-n_0))$ and thus it will be often replaced by its support $\tau=n-n_0$ in the following.

Consider now spatial dynamics. At each iteration, the particle might jump to one of the nearest neighboring sites of the lattice, or stay at rest. The actual local trajectory is not deterministic, i.e., it is not a prescribed function of previous parts of trajectory. Rather, the local trajectory has the characteristics of a random walk. This point is very important and it implies that an intrinsic randomness affects the particle motion. Generally, there is a different transition probability for each of the three possible transitions. In free motion without interference or external forces, these probabilities do not change with time. Let me denote the transition probabilities $a:=\Pr(\mathcal{V}_n=1)$, $b:=\Pr(\mathcal{V}_n=0)$, and $c:=\Pr(\mathcal{V}_n=-1)$, respectively. Of course, 
\begin{equation}
  a+b+c=1.
  \label{eqn:abc}
\end{equation}
Moreover, the proposed model assumes that the expected value of $\mathcal{V}_n$ is imprinted to the particle. This imprint is to be attributed to the preparation and is actualized every time the particle interacts with the environment (in a way to be considered later). Let me denote this expected value as \emph{momentum propensity} $p$,
\begin{equation}
  p := E[\mathcal{V}_n]=a-c.
  \label{eqn:p}
\end{equation}

Another characteristic of the random motion is the expected value of the squared velocity, that is,
\begin{equation}
  e := E[\mathcal{V}_n^2] = a+c
  \label{eqn:e}
\end{equation}
that can be reinterpreted as an \emph{energy propensity}. %(define the stochastic variable ``energy" as $\mathcal{E}_n:=\mathcal{V}_n^2$). 
Combining (\ref{eqn:abc})--(\ref{eqn:e}), obtain
\begin{equation}
  a = \frac{e+p}{2}, \quad b = 1-e, \quad c = \frac{e-p}{2}.
  \label{eqn:a}
\end{equation}

The energy $e$ must be a function of $p$. A well-known result of special relativity states that energy of a particle is the sum of the rest energy and the kinetic energy. Following this suggestion, the proposed model assumes that 
\begin{equation}
  e(p) := \frac{1+p^2}{2}.
  \label{eqn:energy}
\end{equation}
Equation (\ref{eqn:energy}) might be also interpreted in the following way: energy is the average of the ``time energy" and the ``space energy", where the former contribution is always one, since time can only advance by one unit.
Consequently, (\ref{eqn:a}) can be rewritten as
\begin{equation}
  a = \left(\frac{1+p}{2}\right)^2, \quad b = \frac{1-p^2}{2}, \quad c = \left(\frac{1-p}{2}\right)^2.
  \label{eqn:a1}
\end{equation}
Appendix~\ref{app:matter} shows that it is possible to retrieve the de Broglie relation $E=\hbar \omega$ for ``matter waves" with the proposed model, involving the energy propensity $e$.

\subsubsection{Probability mass functions}
%%%%%%%%%%%%%%%%%%%%%%%%%%%%%%%%%%%%%%%%%%%%%%%%%
The equations in the previous section, in fixing the probability of each jump at each iteration $n$, define the trajectory of the particle as a random walk. Let me introduce the stochastic variables that describe the position of the particle, $\mathcal{X}_n$,
\begin{equation}
  \mathcal{X}_n := \xi_0+\sum_{n'=n_0}^{n} \mathcal{V}_{n'}.
	\label{eqn:xstoc}
\end{equation}
where $\xi_0$ denotes the site occupied at $n=n_0$, i.e., $\tau=0$.

Let me calculate the \emph{probablity mass function} of this stochastic variable, 
\begin{equation}
  \rho_{n}(\xi):=\Pr(\mathcal{X}_{n}=\xi), 
  \label{eqn:rho}
\end{equation}
i.e., the probability that the particle has crossed $\xi$ sites after $n$ iterations. Consider the scenario where particles are emitted from a source located at the site $\xi_0=0$ of the lattice with an intrinsic value of $p$, and thus of $e$, determined by the preparation. Time interval between two emissions is very large, so to exclude any interactions between successive particles. Moreover, the single source excludes quantum interference. After one iteration or, equivalently, time step, the particle has a probability $a$ to be at the site $\xi=1$, a probability $b$ to be at the site $\xi=0$, and a probability $c$ to be at the site $\xi=-1$. After two iterations, the probabilities are: $\rho_2(2)=a^2$, $\rho_2(1)=2ab$, $\rho_2(0)=2ac+b^2$, $\rho_2(-1)=2bc$, $\rho_2(-2)=c^2$. Note that, since the functions $a(p)$ and $c(p)$ are symmetric, the probability function is symmetric with respect to $\xi=0$.

In general, the position probability function is described by the recursive equation
\begin{equation}
  \rho_{n}(\xi) = a\rho_{n-1}(\xi-1)+b\rho_{n-1}(\xi)+c\rho_{n-1}(\xi+1).
  \label{eqn:process}
\end{equation}

Deriving a closed formula for the probability mass function $\rho_{n}(\xi)$ is tedious but straightforward at this point. With $n_0=0$ and the initial condition $\rho_0(0)=1$, the result is
\begin{equation}
  \rho_{n}(\xi) = 
	\displaystyle\frac{\left(
\begin{array}{c}
	2\tau \\ \tau+\xi
\end{array}
 \right)}{2^{2\tau}}(1+p)^{\tau+\xi}(1-p)^{\tau-\xi},
  \label{eqn:rhop}
\end{equation}
with support $\xi=\{-\tau,\ldots,\tau\}$, and can be verified by inspection. From this formula, the probability that a particle is at the event horizon, i.e., $\rho_{n}(\tau)$, is easily retrieved as $a^\tau$. Similarly, $\rho_{n}(-\tau)=c^\tau$. The estension to the more general case of $\xi_0\neq 0$ is easily done by replacing $\xi$ with $\xi-\xi_0$ in the right-hand side of (\ref{eqn:rhop}). Finally, it turns out that $E[\mathcal{X}_n]=p\tau+\xi_0$.

The function (\ref{eqn:rhop}) has a limit for large $\tau$'s that can be derived in two equivalent ways. On the one hand, the equation of motion (\ref{eqn:xstoc}) can be reviewed in the continuum limit as a stochastic differential equation reading
\begin{equation}
  d\mathcal{X}_{n} = E[\mathcal{V}_{n}]dn + \sqrt{\mathrm{Var}[\mathcal{V}_{n}]}dB,
	\label{eqn:sde}
\end{equation}
where $dB$ is a Brownian motion with zero mean and unit variance. Now, from (\ref{eqn:p})--(\ref{eqn:a1}), the identity $\mathrm{Var}[\mathcal{V}_{n}]=e-p^2=b$ follows. Consequently, the continuum limit of $\rho$ is a Gaussian function with mean $p\tau$ and variance $b\tau$, that is,
\begin{equation}
  \rho(\xi,\tau) := \lim_{\tau\rightarrow\infty} \rho_{n}(\xi) \approx \frac{1}{\sqrt{2\pi b \tau}} \exp\left(-\frac{(\xi-p\tau)^2}{2b\tau}\right).
 \label{eqn:normal}
\end{equation}
Other derivations of (\ref{eqn:normal}) are illustrated in Appendix~\ref{app:normal}. Appendix~\ref{app:specrel} shows that (\ref{eqn:normal}) is invariant to Lorentz transformations.

Let me define an additional stochastic variable for later use, representing the distance covered by the particle
\begin{equation}
  \mathcal{L}_{n}:=\sum_{n'=n_0}^{n} \mathcal{V}_{n'} = \mathcal{X}_n-\xi_0.
	\label{eqn:Lstoc}
\end{equation}
Clearly, $\Pr(\mathcal{L}_n=\lambda)=\Pr(\mathcal{X}_n=\xi_0+\lambda)$ and $E[\mathcal{L}_n]=p\tau$ hold. 
%The second stochastic variable is the sample momentum
%%
%\begin{equation}
  %\mathcal{Q}_{n}:=\frac{1}{\tau}\sum_{n'=n_0}^{n} \mathcal{V}_{n'} = \frac{\mathcal{L}_n}{\tau}.
	%\label{eqn:Qstoc}
%\end{equation}
%%
%for which $\Pr(\mathcal{Q}_n=q)=\Pr(\mathcal{L}_n=q\tau)$ and $E[\mathcal{Q}_n]=p$ hold.

%%%%%%%%%%%%%%%%%%%%%%%%%%%%%%%%%%%%%%%%%%%%%%%%
\subsection{Lattice dynamics}
%%%%%%%%%%%%%%%%%%%%%%%%%%%%%%%%%%%%%%%%%%%%%%%%%

To describe particle dynamics as ``seen" by the lattice sites, let me define additional stochastic variables. Denote these variables with a time subscript \emph{and} a position superscript, instead of the only time subscript as for the particle variables. 
%In principle, any particle stochastic variable can be transformed into a lattice stochastic variable by imposing that $\sum_0^{\tau} \upsilon_{\tau'}=\xi$. 
Clearly, the variable $\mathcal{X}_{\tau}^{\xi}$ can only take the value $\xi$ in the single-source scenario under consideration. Similarly, a variable $\mathcal{T}_{\tau}^{\xi}$ would be deterministic, with its support including only the value $\tau$.

Let me introduce a true stochastic variable, the \emph{site occupancy}, defined as
\begin{equation}
  \mathcal{O}_{\tau}^{\xi} = \left\{\begin{array}{ll} 1, & \text{if the particle occupies the site $\{\xi,\tau\}$} \\
		0, & \text{otherwise} \end{array} \right.
\end{equation}
Define $\rho_{\tau}^{\xi}:=\Pr(\mathcal{O}_{\tau}^{\xi}=1)=E[\mathcal{O}_{\tau}^{\xi}]$. Clearly,
\begin{equation}
  \rho_{\tau}^{\xi}=\rho_{\tau}(\xi)
\end{equation}
and its pmf is given by (\ref{eqn:rhop}).

%%%%%%%%%%%%%%%%%%%%%%%%%%%%%%%%%%%%%%%%%%%%%%%%%
\subsection{Ensembles of Particles}
%%%%%%%%%%%%%%%%%%%%%%%%%%%%%%%%%%%%%%%%%%%%%%%%%

In this section the predictions of QM are retrieved for ensembles of similarly prepared particles, that is, particles emitted from the same source at $\xi_0=0$.

\subsubsection{Probability density}  \label{sec:wave}
%%%%%%%%%%%%%%%%%%%%%%%%%%%%%%%%%%%%%%%%%%%%%%%%%
To retrieve the predictions of Schr\"{o}dinger's equation, a key element of the model is introduced. The proposed model assumes that whenever the particle interacts with the environment, its momentum propensity $p$ is properly reset. Now, in the free motion scenario, consider for the moment that $p$ is a continous variable determined \emph{randomly} during the preparation at the particle source. Consequently, the probability of releasing a particle with a momentum propensity $p$ is uniform over the interval between -1 and +1, spannnig 2, and thus the probability density of the momentum propensity is $f(p)=1/2$.

When the source releases a large number of particles in succession, each one with a randomly determined value of $p$, the probability of finding a particle at the location $\{\xi,\tau\}$ is given by the \emph{ensemble average} and is calculated as
\begin{equation}
  P_{\tau}^{\xi} := \int_{-1}^{1}f(p)\rho_{\tau}^{\xi}dp = \frac{1}{2}\int_{-1}^{1}\rho_{\tau}^{\xi}dp.
  \label{eqn:intp}
\end{equation}

Introducing (\ref{eqn:rhop}) into (\ref{eqn:intp}), and after some manipulations (see Appendix~\ref{app:P}), obtain
\begin{equation}
  P_{\tau}^{\xi} = \frac{1}{2\tau+1}, \forall \xi \in [-\tau,\tau].
  \label{eqn:P}
\end{equation}
Moreover, it is easily verified that 
\begin{equation}
  \sum_{\xi=-\tau}^{\xi=\tau} P_{\tau}^{\xi} = 1
  \label{eqn:sumrho}
\end{equation}
as obviously required.

Now, compare this result with the predictions of QM, i.e., the particular solution of the Schr\"{o}dinger equation. For a single perfectly localized source at $x=0$, the probability density is calculated (see \ref{app:single}) as
\begin{equation}
  |\Psi(x,t)|^2 = \frac{mX^2}{2\pi\hbar t} = \frac{mX^2}{ht},
  \label{eqn:f0}
\end{equation}
thus it is inversely proportional to time and it does not depend on $x$. Normalizing to lattice units and using (\ref{eqn:hbar}) allows reducing (\ref{eqn:f0}) to
\begin{equation}
  |\Psi(\xi,\tau)|^2 = \frac{1}{2\tau}.
  \label{eqn:f1}
\end{equation}
This result compares with (\ref{eqn:P}), with $2\tau$ replacing $2\tau+1$. The two functions of $\tau$ are very similar and, indeed, practically coincident for $\tau$ sufficiently large. In other terms, \emph{the square modulus of the wavefunction predicted by the Schr\"{o}dinger equation is the continuum limit of the probability in the proposed model}. 

Note that only the particular formulation of energy propensity (\ref{eqn:energy}) yields this result. Other values for $e$ and consequently for $b$ (for instance, $b\equiv 0$), would yield position-dependent mass probability functions.

\subsubsection{Phase} \label{sec:phase}
%%%%%%%%%%%%%%%%%%%%%%%%%%%%%%%%%%%%%%%%%%%%%%%%%

To retrieve the phase of the wavefunction predicted by the Schr\"{o}dinger equation, define first a new lattice stochastic variable $\mathcal{S}_{\tau}^{\xi}$ representing the accumulated energy of the particle as seen by a site. Its pmf is calculated (see Appendix~\ref{app:phis}) as 
\begin{equation}
  \phi_{\tau}^{\xi}:=\Pr(\mathcal{S}_{\tau}^{\xi}=\sigma) = 
N_p(n_c)\frac{2^{n_b}}{{2\tau \choose \tau+\xi}} = \displaystyle\frac{2^{\tau-\sigma} \cdot {\tau \choose \frac{\sigma+\xi}{2}} \cdot {\tau-\frac{\sigma+\xi}{2} \choose \tau-\sigma}}{{2\tau \choose \tau+\xi}},
  \label{eqn:PrS}
\end{equation}
with support $\sigma=\{|\xi|,|\xi|+2,\ldots,|\xi|+2\left\lfloor \frac{\tau-|\xi|}{2} \right\rfloor\}$.
It can be also proved that
\begin{equation}
  E[\mathcal{S}_{\tau}^{\xi}] = |\xi|+\frac{(|\xi|-\tau)(|\xi|-\tau+1)}{(2\tau-1)} = \frac{\xi^2+\tau^2-\tau}{(2\tau-1)}
	\label{eqn:eS}
\end{equation}
and
\begin{equation}
  \mathrm{Var}[\mathcal{S}_{\tau}^{\xi}] = 2\frac{(\xi^2-\tau^2)(\xi^2-(\tau-1)^2)}{(2\tau-1)^2(2\tau-3)}.
	\label{eqn:VarL}
\end{equation}
Equation (\ref{eqn:eS}) can be easily verified by inspection of a few sites, provided that $\{\xi,\tau\}>0$. For example, for $\xi_0=0$, one easily obtains $E[\mathcal{S}_{1}^{1}]=1$. The cumulated energy of every particle reaching the site $\{1,1\}$ is obviously 1. Generalizing this result, clearly $E[\mathcal{S}_{\tau}^{\tau}]=\tau$. For a site like $\{0,2\}$ the prediction is less trivial. The possible values of the cumulated energy can be 2, if the particle follows a back-and-forth path, or 0, if it stays at rest for two time steps. Using (\ref{eqn:eS}), one obtains for this case $E[\mathcal{S}_{2}^{0}]=2/3$, which is a weighted mean between the two possible values of cumulated energy. Equation (\ref{eqn:VarL}) also can be verified by inspection. For instance, when $\xi=\pm\tau$ or $\xi=\pm(\tau-1)$, it is apparent that $\mathcal{S}_{\tau}^{\xi}$ can only take the value $|\xi|$. In fact, (\ref{eqn:VarL}) yields $\mathrm{Var}[\mathcal{S}_{\tau}^{\xi}] =0$. 

The continuum limit counterpart of (\ref{eqn:PrS}) is calculated as
\begin{equation}
  {\phi}(\sigma;\xi,\tau) := \lim_{\tau\rightarrow\infty} \phi_{\tau}^{\xi} = \frac{1}{\sqrt{2\pi \mathrm{Var}[\mathcal{S}_{\tau}^{\xi}]}} \exp\left(- \frac{(\sigma-E[\mathcal{S_{\tau}^{\xi}}])^2}{2\mathrm{Var}[\mathcal{S}_{\tau}^{\xi}]}\right)
	\label{eqn:phis}
\end{equation}
Define now the \emph{action} at any site as
\begin{equation}
  \Sigma_{\tau}^{\xi}:=\int_0^{\tau} f(p) E[\mathcal{S}_{\tau}^{\xi}] dp.
	\label{eqn:action}
\end{equation}
 %as the expected value of the stochastic variable obtained by summing the values of the average energy along the path, i.e., $\mathcal{S}_{\tau}^{\xi}:=\sum_0^{\tau}\mathcal{E}_{\tau'}^{\xi_{\tau'}}$. This variable can be also regarded as the cumulated energy possessed by the particle when it passes through a certain site at a certain time, $\mathcal{S}_{\tau}^{\xi}=\sum_0^{\tau}|\beta_{\tau'}|$, s.t. $\sum_0^{\tau} \beta_{\tau'}=\xi$.
%
%The following recursive equation is easily derived,
%%
%\begin{equation}%\begin{split}
  %\Sigma_{\tau}^{\xi} = \frac{1}{\rho_{\tau}^{\xi}}\left(a\rho_{\tau-1}^{\xi-1}(\Sigma_{\tau-1}^{\xi-1}+1) %+\right. \\ & 
	%+b\rho_{\tau-1}^{\xi}\Sigma_{\tau-1}^{\xi}+%\left.
	%c\rho_{\tau-1}^{\xi+1}(\Sigma_{\tau-1}^{\xi+1}+1)\right),
  %\label{eqn:sigma}%\end{split}
%\end{equation}
%%
%with the initial condition $\Sigma_0^{\xi_0}=0$.
%Combining (\ref{eqn:a1})--(\ref{eqn:rhop}) with (\ref{eqn:sigma}) yields
Note that the expected value given by equation (\ref{eqn:eS}) does not depend on $p$. Therefore, by inserting it into (\ref{eqn:action}), obtain
\begin{equation}
  \Sigma_{\tau}^{\xi} = \frac{\xi^2+\tau^2-\tau}{2\tau-1} = \Sigma_{\tau}^{\xi_0}+\frac{\xi^2}{2\tau-1}.
  \label{eqn:s}
\end{equation}
%
%Equation (\ref{eqn:s}) can be easily verified by inspection of a few points, provided that $\{\xi,\tau\}>0$. For example, for $\xi_0=0$, one easily obtains $\Sigma_1^1=1$. The action of every particle reaching the point $\{1,1\}$ is obviously 1. Generalizing this result, clearly $\Sigma_{\tau}^{\tau}=\tau$. For a point like $\{0,2\}$ the prediction is less trivial. The possible values of action can be 2, if the particle follows a back-and-forth path, or 0, if it stays at rest for two time steps. Using (\ref{eqn:sigma}), one obtains for this case $\Sigma_2^0=2/3$, which is a weighted mean between the two possible values of action. 

%Note from (\ref{eqn:ebar}) that 
%%
%\begin{equation}
  %\Sigma_{\tau}^{\xi}=\tau\epsilon_{\tau}^{\xi}
%\end{equation}
%%
%as expected from the usual definition of action as the integral of energy. Analogously, $\xi=E[\mathcal{X}_{\tau}^{\xi}]=\tau q_{\tau}^{\xi}$. 

Compare this result with the phase of the wavefunction (\ref{eqn:psi2}). The latter, usually interpreted as the action of the particle, is
\begin{equation}
  S(x,t)=\frac{mx^2}{2\hbar t},
\end{equation}
which, in lattice units, becomes
\begin{equation}
  S(\xi,\tau)=\pi\frac{\xi^2}{2\tau}.
  \label{eqn:S}
\end{equation}
The latter equation corresponds to the second term in the right-hand side of (\ref{eqn:s}), that is, $\Sigma_{\tau}^{\xi}-\Sigma_{\tau}^{\xi_0}$, multiplied by $\pi$ to obtain a phase angle. The correspondence is almost perfect, except for the term $2\tau-1$ that in the proposed model replaces the term $2\tau$ predicted by QM. For large values of $\tau$, however, the two results are practically coincident. In other terms, \emph{the phase of the wavefunction predicted by the complex Schr\"{o}dinger equation is the continuum limit of the action in the proposed model}.

%As a partial result to be used later, let me calculate now the variance of $\mathcal{S}_{\tau}^{\xi}$. Let me introduce the auxiliary stochastic variable $\mathcal{\tilde{S}}_{\tau}^{\xi}:=\displaystyle\frac{\mathcal{S}_{\tau}^{\xi}-|\mathcal{X}_{\tau}^{\xi}|}{2}$. The support values of $\mathcal{\tilde{S}}_{\tau}^{\xi}$ are $\tilde{\sigma}=0,1,\ldots,\sigma_M$, where $\sigma_M:=\left\lfloor \frac{\tau-|\tilde{\xi}|}{2} \right\rfloor$. It can be verified by inspection that
%%
%\begin{equation}
  %E[\mathcal{\tilde{S}}_{\tau}^{\xi}] = \frac{(|\tilde{\xi}|-\tau)(|\tilde{\xi}|-\tau+1)}{2(2\tau-1)}
%\end{equation}
%%
%and
%%
%\begin{equation}
  %\mathrm{Var}[\mathcal{\tilde{S}}_{\tau}^{\xi}] = \frac{(\tilde{\xi}^2-\tau^2)(\tilde{\xi}^2-(\tau-1)^2)}{2(2\tau-1)^2(2\tau-3)}.
	%\label{eqn:VarL}
%\end{equation}
%%
%For instance, when $\tilde{\xi}=\pm\tau$ or $\tilde{\xi}=\pm(\tau-1)$, it is apparent that $\mathcal{S}_{\tau}^{\xi}$ can only take the value $|\tilde{\xi}|$. Consequently, $\sigma_M=0$ and $\mathrm{Var}[\mathcal{\tilde{S}}_{\tau}^{\xi}] =0$ as predicted by (\ref{eqn:VarL}). By virtue of definition, $\Sigma_{\tau}^{\xi}$ is retrieved as $|\tilde{\xi}|+2E[\mathcal{\tilde{S}}_{\tau}^{\xi}]$, yielding (\ref{eqn:s}). On the other hand, $\mathrm{Var}[\mathcal{S}_{\tau}^{\xi}] = 4\mathrm{Var}[\mathcal{\tilde{S}}_{\tau}^{\xi}] $, which is the result sought.

\subsubsection{Schr\"{o}dinger equation: de Broglie-Bohm formulation}
%%%%%%%%%%%%%%%%%%%%%%%%%%%%%%%%%%%%%%%%%%%%%%%%%

The results in the previous sections have been derived for a probability density of the momentum propensity $f(p)=1/2$. Consider now a generic function $f(p)$. Apply (\ref{eqn:intp}) to the continuum limit (\ref{eqn:normal}) of the probability function $\rho_{\tau}^{\xi}$. It can be shown that (\ref{eqn:normal}) approximates a Dirac delta function, whence
\begin{equation}
  {P}(\xi,\tau) := \int_{-1}^1 f(p)\rho(\xi,\tau) dp \approx \frac{f(\xi/\tau)}{\tau}.
  \label{eqn:fsut}
\end{equation}
The probability density function $P$ obeys the following partial differential equation
\begin{equation}
  \frac{\partial P}{\partial \tau} = -\frac{\xi}{\tau} \frac{\partial P}{\partial \xi} - \frac{1}{\tau} P = -\frac{\partial}{\partial \xi} \left( \frac{\xi}{\tau}P\right).
	\label{eqn:schro1}
\end{equation}
Introducing now the continuum-limit approximation of $\Sigma_{\tau}^{\xi}$,
\begin{equation}
  \Sigma(\xi,\tau):=\lim_{\tau\rightarrow\infty} \Sigma_{\tau}^{\xi} \approx \frac{\xi^2+\tau^2}{2\tau},
\end{equation}
one recognizes in (\ref{eqn:schro1}) the continuity equation 
\begin{equation}
  \frac{\partial P}{\partial \tau} = - \frac{\partial}{\partial \xi} \left(P \frac{\partial \Sigma}{\partial \xi} \right).
  \label{eqn:continuity}
\end{equation}
On the other hand, the relationship 
\begin{equation}
  \frac{\partial \Sigma}{\partial \tau} = -\frac{1}{2}\left(\frac{\partial \Sigma}{\partial\xi}\right)^2
  \label{eqn:hjb}
\end{equation}
also holds. 

Sect.~\ref{sec:wave} has shown the equivalence of $P$ to $|\Psi|^2$, while Sect.~\ref{sec:phase} that of $\Sigma$ with $S/\pi=\angle\Psi/\pi$ for large $\tau$'s. With these two substitutions, and reintroducing physical units instead of lattice units, equations~(\ref{eqn:continuity})--(\ref{eqn:hjb}) become the continuity equation
\begin{equation}
  \frac{\partial |\Psi|^2}{\partial t} = - \frac{\partial}{\partial x} \left(|\Psi|^2 \frac{\partial S\hbar/m}{\partial x} \right)
  \label{eqn:continuity2}
\end{equation}
and the Hamilton--Jacobi equation
\begin{equation}
  \frac{\partial S}{\partial t} = -\frac{1}{2m}\left(\frac{\partial S\hbar}{\partial x}\right)^2,
  \label{eqn:hjb2}
\end{equation}
of the de Broglie--Bohm formulation of QM, which in turn are equivalent to Schr\"{o}dinger's equation for a free particle.

%It is also interesting to note that $\partial \sigma/\partial \xi$ approximates $q$ for large $\tau$'s, while $\partial \sigma/\partial \tau\approx -\epsilon$.

%%%%%%%%%%%%%%%%%%%%%%%%%%%%%%%%%%%%%%%%%%%%%%%%%
\subsection{Numerical results}
%%%%%%%%%%%%%%%%%%%%%%%%%%%%%%%%%%%%%%%%%%%%%%%%%
In the last sections, the predictions of the proposed model were shown in closed form, using mathematical equations in terms of \emph{a priori} probabilities and probability fluxes. The probability of a number of observable were calculated and found to be in accord to the predictions of QM. Now, I will present numerical simulations of the random walk of \emph{single} particles and I will calculate the \emph{a posteriori} probabilities as frequencies over a large number of emissions. Thus, this section is aimed at reproducing numerically a true experiment. 

Table~\ref{tab:pseudo} shows the algorithm used for such simulations. The two for-cycles are for the successively released $N_P$ particles, and for time up to $N_T$. Each particle experiences the choice of two randomly-selected values: (i) the momentum propensity and (ii) at each time step, its local velocity $\upsilon$ as a function of $p$. 
%
%Note that for this scenario the term $\gamma$ is identically null. 
%
The final code line represents the counting of the particle that arrive at a certain location at time $N_T$. From this number of arrivals, an \emph{a posteriori} frequency $\nu(\xi)$ is calculated as the ratio to the total number of particles emitted.

\begin{algorithm}
\caption{Pseudocode used for the simulations of Fig.~\ref{fig:buildprob}.} \label{tab:pseudo}
\begin{algorithmic}[1]
\For {particle 1 to $N_P$}
\State $\xi \gets 0$
\State $\tau \gets 0$
\State $p \gets$ random value beween -1 and +1
\For{iteration 1 to $N_T$}
\State $\tau \gets \tau+1$
\State $\upsilon \gets$ random value +1, 0 or -1 with prob. given by (\ref{eqn:a})
\State $\xi \gets \xi+\upsilon$
\EndFor
\State $\nu(\xi) \gets \nu(\xi)+1$ 
\EndFor 
\State $\nu(\xi) = \nu(\xi)/N_P$
\end{algorithmic}
\end{algorithm}

\begin{figure}[t!]
  \centering
  \includegraphics[width=\textwidth]{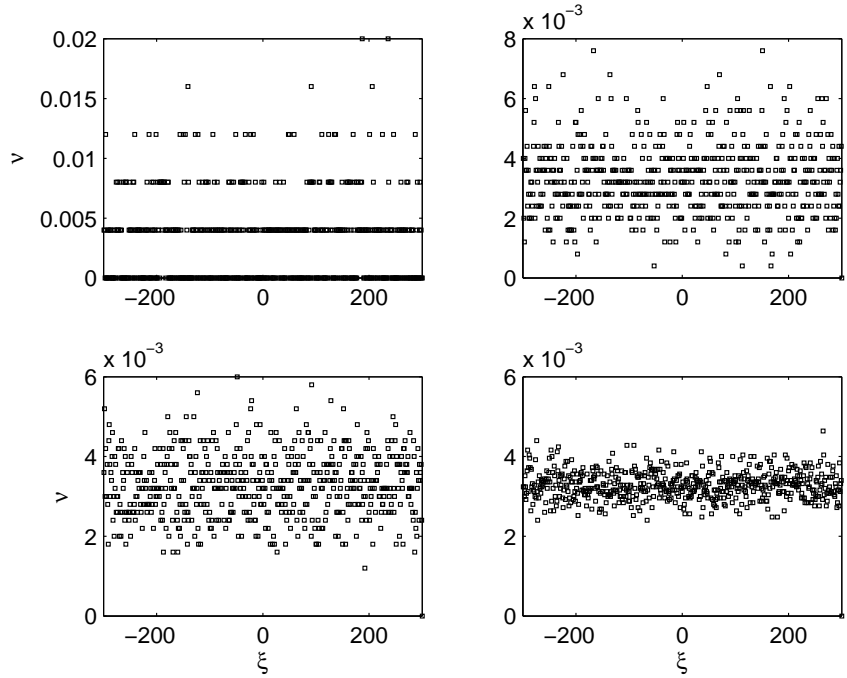}
  \caption{Frequency of arrival of particles emitted at $\xi=0$ as a function of $\xi$ after $N_T=300$. From top-left to bottom-right, $N_P=$ 500, 5000, 10000, and 50000, respectively.}
  \label{fig:buildprob}
\end{figure}

Figure~\ref{fig:buildprob} shows the frequency $\nu(\xi)$ after a time $N_T=300$ for different values of $N_P$. As the the number of particles emitted in the ensemble increases, a frequency distribution builds up. For large $N_P$, the frequency clearly tends to the a priori probability $P(\xi,\tau=N_T)$, that is, a constant value given by (\ref{eqn:P}).

%%%%%%%%%%%%%%%%%%%%%%%%%%%%%%%%%%%%%%%%%%%%%%%%%
%%%%%%%%%%%%%%%%%%%%%%%%%%%%%%%%%%%%%%%%%%%%%%%%%
\section{Interference} \label{sec:interf}
%%%%%%%%%%%%%%%%%%%%%%%%%%%%%%%%%%%%%%%%%%%%%%%%%
%%%%%%%%%%%%%%%%%%%%%%%%%%%%%%%%%%%%%%%%%%%%%%%%%

%%%%%%%%%%%%%%%%%%%%%%%%%%%%%%%%%%%%%%%%%%%%%%%%%
%\subsection{Double-slit preparation} \label{sec:prep}
%%%%%%%%%%%%%%%%%%%%%%%%%%%%%%%%%%%%%%%%%%%%%%%%%
After having reproduced the predictions of the Schr\"{o}dinger equation for a free particle, let me proceed now to a second puzzling aspect of QM: particle self-interference. Double-slit experiment usually serves to visualize this phenomenon. However, the core of self-interference is isolated and better illustrated by a \emph{double-source} preparation, where particles can be emitted by two alternative sources and the two possible paths interefere with each other. Additional free-particle scenarios leading to self-intereference are multiple-source preparations and the ``particle in a ring" situation.

%Instead of having a single source, a two-slit barrier, and a screen behind the barrier, I will represent the same process with two independent and mutually alternative sources of particles, and a screen. The sources are equivalent to very narrow, i.e., punctiform slits. In a one-dimensional space, the location of the ``screen" is clearly fictitious. Pictorially, the geometry of the system can be still imagined in two spatial dimensions. One dimension is $\xi$, along which the particle move with a momentum propensity $p$. The second dimension is perpendicular to $\xi$ and is traversed by the particle with momentum propensity 1 (certainty of advancing in the positive direction). The ``screen" is thus located at a distance $\tau$ from the sources. Figure~\ref{fig:slits} illustrates this equivalence.

%\begin{figure}[t!]
  %\centering
  %\includegraphics[width=.6\textwidth]{slits.eps}
  %\caption{Pictorial equivalence between a double-slit experiment in two dimensions and a double-source test case in one dimension.}
  %\label{fig:slits}
%\end{figure}

The representation of these scenarios using only the process illustrated in Sect.~\ref{sec:single} would give just the superimposition of probability densities of the type (\ref{eqn:P}) and no interference would arise. Instead, interference is originated in the proposed model from the interaction of successive particles emitted with the lattice, a process denoted here as ``quantum force," which is illustrated in the next sections.

%%%%%%%%%%%%%%%%%%%%%%%%%%%%%%%%%%%%%%%%%%%%%%%%%
\subsection{Quantum force mechanism} \label{sec:qforce}
%%%%%%%%%%%%%%%%%%%%%%%%%%%%%%%%%%%%%%%%%%%%%%%%%

In short, the emergence of quantum forces is the result of an exchange of information between the lattice and the particles. Let me call this piece of information \emph{boson}, in analogy with force-mediating particles.

In the proposed model, each particle carries two counters. The spatial counter records the number of space steps crossed and thus it is represented by the stochastic variable $\mathcal{L}_{n}$. The temporal counter records the number of time steps crossed and thus is always certainly equal to $\tau$. 

On the other hand, each lattice site maintains a sort of ``register" storing the value of the spatial counter of the last particle that has visited the site. Let me describe this register by the lattice stochastic variable $\mathcal{L}_{\tau}^{\xi}$. 

When a particle visits a site, the site senses the value of the particle's spatial counter, say, $\lambda$. This value is compared with the value $\mu$ stored in the register, which is the value sensed the last time a particle visited the site. If the difference $\delta:=|\mu-\lambda|$ is strictly positive, an exchange of information between the particle and the lattice site takes place.

First, a couple of ``bosons" is created. One boson is taken by the particle (later it will be explained how this boson acts on the particle), while the second stays at the site. The pair $\{\lambda,\mu\}$ for which these bosons have been created distinguishes them from other possible bosons. For this reason let me call them $\lambda\mu$-bosons. 

The second event taking place when $\delta>0$ is that the value $\mu$ in the site register is replaced with the value $\lambda$, while the particle counter is replaced with the value $\mu$. In other words, the site register and the particle counter are exchanged,
\begin{equation}
  \mathcal{L}_{\tau}^{\xi} \leftarrow \lambda, \quad \mathcal{L}_{n} \leftarrow \mu.
\end{equation}

The next subsections will describe separately the behavior of site bosons and particle bosons.

\subsubsection{Quantum force at the lattice}
%%%%%%%%%%%%%%%%%%%%%%%%%%%%%%%%%%%%%%%%%%%%%%%%%

When a new $\lambda\mu$-boson is created at a lattice site, first it suppresses the possibly resident boson of the same type, since the site can carry multiple bosons only if they are characterized by different values $\{\lambda,\mu\}$. Then the new boson is assigned a momentum $w_{\lambda\mu}^{(0)}$ that equals the sample momentum of the particle. If $\mathcal{W}_{\xi,\lambda\mu}^{\tau}$ is the stochastic variable describing the site boson momentum, a re-initialization of such variable occurs whenever a new boson pair is created according to the rule
\begin{equation}
  \mathcal{W}_{\xi,\lambda\mu}^{\tau} \leftarrow w_{\lambda\mu}^{(0)} = \frac{\lambda}{\tau}.
  \label{eqn:wboson0}
\end{equation}
One can interprete this rule as an exchange of momentum between the site and the particle (considering its \emph{sample} momentum), through their respective bosons.

When the site is not visited or it senses a particle counter $\lambda=\mu$ ($\delta=0$), the momentum of the resident $\lambda\mu$-boson decays according to the rule
%The value of this momentum changes as more particles visit the site. Consider the $\ell$-th boson of the same species (i.e., a $\lambda\mu$-boson) created by the site. Its momentum reads
%
\begin{equation}
	w_{\lambda\mu}^{(\ell)} = w_{\lambda\mu}^{(\ell-1)}\left(1-\frac{(\delta w_{\lambda\mu}^{(0)})^2}{\ell^2} \right), \quad \ell=1,2,\ldots,
	\label{eqn:wdecay}
\end{equation}
where $\ell$ is the lifetime of the resident boson, that is, the number of iterations spanned from when it was created ($\ell=0$ for a new boson). 

Rule (\ref{eqn:wdecay}) applies separately to each boson type. Note that this rule may be interpreted as a discrete analogous of an exponential decay. Note also that, since $\lambda/\tau \in \mathbb{Q}$ and $\delta, \ell \in\mathbb{N}$, also $w_{\lambda\mu}^{(\ell)}, p_{\lambda\mu}^{(0)} \in \mathbb{Q}$.

\subsubsection{Quantum force on the particle}
%%%%%%%%%%%%%%%%%%%%%%%%%%%%%%%%%%%%%%%%%%%%%%%%%

Similarly to site bosons, when a new particle $\lambda\mu$-boson is created, it suppresses the possibly carried boson of the same type, as the particle can carry multiple bosons only if they are characterized by different values $\{\lambda,\mu\}$. 

%Each time a particle visits a site, it takes a new $\lambda\mu$-boson if the value of its spatial counter, $\lambda$, is different from the value $\mu$ of the site register. If a new boson is not taken, the particle keeps the boson it has, if it has one. 
%Therefore, at each time, the particle configuration regarding a possible $\lmbda\mu$-boson might be one of the three following cases:
%%
%\begin{enumerate}
	%\item The particle does not carry a $\lambda\mu$-boson, that is, it has never visited a site having $\mathcal{X}_{\tau}^{\xi}=\mu$ with $\mathcal{X}_{\tau}=\lambda$;
	%\item The particle carries a ``new" $\lambda\mu$-boson, that is, it has just visited a site having $\mathcal{X}_{\tau}^{\xi}=\mu$ with $\mathcal{X}_{\tau}=\lambda$;
	%\item The particle carries an ``old" $\lambda\mu$-boson, that is, it has not just visited a site having $\mathcal{X}_{\tau}^{\xi}=\mu$ with $\mathcal{X}_{\tau}=\lambda$, but it has done it in the past.
%\end{enumerate}
%%

The particle boson momentum is assigned a value $p_{\lambda\mu}^{(0)}$ that equals that of the boson of the same type that was previously resident at the site, if there was one. If the momentum of the $\lambda\mu$ boson possessed by the particle is represented by the new stochastic variable $\mathcal{P}_{\lambda\mu}$ (the iteration index is omitted for simplicity), a re-initialization occurs whenever a new boson pair is created, according to the rule
\begin{equation}
  \mathcal{P}_{\lambda\mu} \leftarrow p_{\lambda\mu}^{(0)} = w_{\lambda\mu}^{(\ell)}
	\label{eqn:pboson0}
\end{equation}

When the particle visits a site where new $\lambda\mu$-bosons are not created, the momentum of its $\lambda\mu$-boson decays according to the rule
\begin{equation}
	p_{\lambda\mu}^{(k)} = p_{\lambda\mu}^{(k-1)}\left(1-\frac{1}{2k}\right), \quad k=1,2,\ldots %\beta_{k}, k=0,1,\ldots
	\label{eqn:pdecay}
\end{equation}
where $k$ is the lifetime of the particle boson, that is, the number of iterations spanned from when it was created ($k=0$ for a new boson). Rule (\ref{eqn:pdecay}) may be interpreted as a discrete analogous of a decay that is $\sim 1/\sqrt{k}$.

%and $\beta_{k}$ is a decay term, that varies during the boson lifetime. Referring to the three cases above,
%%
%\begin{enumerate}
	%\item If the particle does not carry a boson, then there is no momentum exchange;
	%\item If the particle carries a ``new" $\lambda\mu$-boson, then $\beta_{0}=1$;
	%\item If the particle carries an ``old" $\lambda\mu$-boson, then
%\begin{equation}
  %\beta_{k}=\beta_{k-1}\left(1-\frac{1}{2k}\right).
	%\label{eqn:beta}
%\end{equation}
%\end{enumerate}

Finally, the effective momentum propensity that regulates the particle motion is not the particle momentum $p$ but its total momentum
\begin{equation}
	\mP := p-\sum_{\lambda,\mu} \mathcal{P}_{\lambda\mu},
	\label{eqn:peff}
\end{equation}
where the summation is taken over all bosons carried by the particle (the $k$ index has been removed from the right-hand side terms for simplicity since, at a certain time step, each boson has its own lifetime). 

Simple as it is, the mechanism given by (\ref{eqn:pboson0})--(\ref{eqn:peff}) is capable of accounting for self-interference, as it will be proven in next section for a number of increasingly complex situations.

%%%%%%%%%%%%%%%%%%%%%%%%%%%%%%%%%%%%%%%%%%%%%%%%%
\subsection{Ensembles of Particles} \label{sec:apriori}
%%%%%%%%%%%%%%%%%%%%%%%%%%%%%%%%%%%%%%%%%%%%%%%%%

In this section the predictions of QM are retrieved for ensembles of similarly prepared particles. The scenarios considered are (i) the two-slit or two-source scenario, (ii) the multiple-source scenario, (iii) the particle-in-a-ring scenario, and 'iv) the particle-in-a-box scenario.

\subsubsection{Two slits}
%%%%%%%%%%%%%%%%%%%%%%%%%%%%%%%%%%%%%%%%%%%%%%%%%

With the mechanism illustrated in Sect.~\ref{sec:qforce}, derive now the probability function $P(\xi,\tau)$ for the two-source scenario, that is, two lattice sources separated by $\delta$ sites, labelled 1 (source at $\delta/2$) and 2 (source at $-\delta/2$), respectively. The probability that a particle is emitted from source 1 is $P_1$, while $P_2=1-P_1$ is the probability of emission from source 2.

In this scenario, any site register $\mathcal{L}_{\tau}^{\xi}$ can take only the values $\xi-\delta/2$ (particle coming from source 1) and $\xi+\delta/2$ (particle coming from source 2). Let me label these two alternative events ``1" and ``2" in the following. Four types of $\lambda\mu$ events are possible, namely, ``11", ``12", ``21", and ``22". The probabilities of these events are easily calculated as $P_{11}:=P_1^2$, $P_{12}:=P_1P_2$, $P_{21}:=P_1P_2$, $P_{22}:=P_2^2$. The events ``11" and ``22" do not create any boson (the particle counter is equal to the site register). Consider now the bosons ``12". 

The boson momentum at a site $\{\xi,\tau\}$ is a stochastic variable whose expected value is calculated in Appendix~\ref{app:W}. The final result is
\begin{equation}
  E[\mathcal{W}_{\xi,12}^{\tau}] \approx \frac{\sin(\pi \delta \frac{\xi}{\tau})}{\pi\delta },
\label{eqn:sinc}
\end{equation}
a result remarkable due to the emergence of a trigonometric function from the discrete-valued model proposed.

%To compute the boson momentum at a site $\{\xi,\tau\}$, apply rule (\ref{eqn:wdecay}) to find
%%
%\begin{equation}
  %w_{12}^{(\ell)} = q_{1} \prod_{j=1}^{\ell} \left(1-\frac{(\delta q_1)^2}{j^2} \right),
	%\label{eqn:product}
%\end{equation}
%%
%where $q_1:=\left(\frac{\xi-\delta/2}{\tau} \right)$. For large boson lifetimes $\ell$, the product in (\ref{eqn:product}) tends to the Sinc of $\delta q_1$, that is
%%
%\begin{equation}
  %w_{12} \approx q_1 {\rm Sinc}(\delta q_1)=\frac{\sin(\pi \delta q_1)}{\pi\delta }.
	%\label{eqn:sinc}
%\end{equation}
%%
%
%The probability that a new site boson is created is the product of three terms: $\rho_{\xi}^{\tau}$ (the probability that the site is visited), the frequency with which particles in the ensemble are emitted, and $P_{12}$. Such probability is so small that the average value of $\mathcal{W}_{\xi,\lambda\mu}^{\tau}$ practically coincides after a certain time (denoted here as lattice ``training" time) with the steady-state value $w_{12}$ above. Moreover, $q_1$ tends to $\xi/\tau$ for large $\xi$ and $\tau$, such that
%%
%\begin{equation}
  %E[\mathcal{W}_{\xi,\lambda\mu}^{\tau}] \approx \frac{\sin(\pi \delta \frac{\xi}{\tau})}{\pi\delta }.
%\label{eqn:Wxt}
%\end{equation}
%%

On the other hand, the expected value of the particle boson momentum is calculated in Appendix~\ref{app:Pt} as
\begin{equation}
  E[\mathcal{P}_{12}] = \sqrt{P_1P_2} \cdot \frac{\sin(\pi \delta \frac{\mathcal{X}_{\tau}}{\tau})}{\pi\delta }.
	\label{eqn:Ep12}
\end{equation}
Again, the result is remarkable since a square-root function emerges from the discrete-valued model proposed.

Finally, the expected value of the particle total momentum is obtained from (\ref{eqn:peff}) as
%
%\begin{equation}
  %E[\mathcal{P}_{12}] = p_{12}^{(0)} \sqrt{P_{12}} = p_{12}^{(0)} \sqrt{P_1P_2}.
	%\label{eqn:pbosmean}
%\end{equation}
%
%and, with the same notation,
%
\begin{equation}
  E[\mP] = p-2 \sqrt{P_1P_2} \cdot \frac{\sin(\pi \delta \frac{\mathcal{X}_{\tau}}{\tau})}{\pi\delta }.
	\label{eqn:Ept}
\end{equation}
since bosons ``12" and ``21" give the same contribution to the combined momentum (while the events ``11" and ``22" do not generate bosons).

The average particle motion is thus described for large $\tau$'s by a couple of equations involving $\bar{\xi}_{\tau}:=\lim_{\tau\rightarrow \infty} E[\mathcal{X}_{\tau}]$ and $\bp:=\lim_{\tau\rightarrow \infty} E[\mP]$, namely,
\begin{equation}
  \bar{\xi}_{\tau+1} = \bx+\bp
	\label{eqn:Ex}
\end{equation}
and
\begin{equation}
  \bp = p - 2\sqrt{P_1 P_2}\cdot \frac{\sin\left(\pi\delta \displaystyle\frac{\bar{\xi}_{\tau}}{\tau}\right)}{\pi \delta}.
	\label{eqn:Ep'}
%	\mathcal{P}_t = p-2\sqrt{P_1 P_2}\cdot \frac{\sin\left(\pi\delta \displaystyle\frac{\mathcal{X}_{\tau}}{\tau}\right)}{\pi \delta} + (\ldots)W,
\end{equation}
%

%Defining $q:=E[\mathcal{P}_t]$, (\ref{eqn:Ep1}) yields
%%
%\begin{equation}
  %q \rightarrow p-2\sqrt{P_1P_2} \cdot \frac{{\rm sin}(\pi\delta q)}{\pi\delta}.
	%\label{eqn:Ep'}
%\end{equation}
%

Using the properties $E[\mathcal{X}_{\tau}]\approx E[\mathcal{L}_{\tau}]=\tau E[\mP]$, the steady-state probability density $f(\bp)$ can be calculated given the probability density of $p$, $f(p)=1/2$. Using (\ref{eqn:Ep'}) and the rule $f(\bp)d\bp=f(p)dp$, obtain
\begin{equation}
  f(\bp) = \frac{1}{2} \frac{dp}{d\bp} = \frac{1}{2} (1+2\sqrt{P_1P_2}\cos(\pi\delta \bp)).
\label{eqn:fq}
\end{equation}
The probability density function $P(\xi,\tau)$ still obeys the approximated rule (\ref{eqn:fsut}), that is
\begin{equation}
  P(\xi,\tau) = \frac{f(\bp=\xi/\tau)}{\tau}.
\end{equation}
From (\ref{eqn:fq}) obtain finally
\begin{equation}
  P(\xi,\tau) = \frac{1+2\sqrt{P_1P_2}\cos\left(\pi\delta \displaystyle\frac{\xi}{\tau}\right)}{2\tau}.
\end{equation}
This result is in perfect agreement with QM predictions, derived in Appendix~\ref{app:2slit} from Schr\"{o}dinger's equation, under the equivalence rule $P \approx |\Psi|^2$ derived in Sect.~\ref{sec:single} and after transformation to lattice units.
%precisely equation (\ref{eqn:p2noneq}).

\subsubsection{Multiple slits}
%%%%%%%%%%%%%%%%%%%%%%%%%%%%%%%%%%%%%%%%%%%%%%%%%

Similar considerations apply for scenarios with $N_s>2$ sources, where $P_i$ is the probability of the i-th source ($\sum_{i=1}^{N_s}P_i=1$) and $\delta_{ij}$ is the distance in lattice units between sources $i$ and $j$, for $i\neq j$. In these cases, there are $N_s(N_s-1)$ possible types of bosons, each of which ultimately contributes to a term $\sqrt{P_iP_j}\cos\left(\pi|\delta_{ij}|\frac{\xi}{\tau}\right)$ to the momentum pdf. Since contribution of boson $ij$ is the same as that of boson $ji$, they can be lumped to yield
\begin{equation}
  f(\bp) = \frac{1+ \sum_{i=1}^{N_s} \sum_{j=i+1}^{N_s} 2\sqrt{P_iP_j}\cos\left(\pi |\delta_{ij}| \bp\right)}{2},
\end{equation}
\begin{equation}
  P(\xi,\tau) = \frac{1+ \sum_{i=1}^{N_s} \sum_{j=i+1}^{N_s} 2\sqrt{P_iP_j}\cos\left(\pi |\delta_{ij}| \displaystyle\frac{\xi}{\tau}\right)}{2\tau}.
	\label{eqn:p2mult}
\end{equation}
Again, this result is in perfect agreement with QM predictions, derived in Appendix~\ref{app:multi} from Schr\"{o}dinger's equation, after transformation to lattice units.

\subsubsection{Particle in a Ring}
%%%%%%%%%%%%%%%%%%%%%%%%%%%%%%%%%%%%%%%%%%%%%%%%%

Also the well-known ``lattice in a ring" scenario can be simulated with the presented model, since the particle has no interactions with the environment. Now the useful lattice does not extend infinitely in both directions but only to values $\xi \in [0,\ell-1]$, where $\ell$ is the ring circumference in lattice units. Due to ring periodicity, it is now necessary to distinguish the stochastic variable $\mathcal{X}_{\tau}$, the site reached after $\tau$ time steps, from the particle spatial counter $\mathcal{L}_{\tau}$ defined by (\ref{eqn:xstoc}). In fact (assume for simplicity $\mathcal{X}_0 \equiv 0$), $\mathcal{X}_{\tau} = \mathcal{L}_{\tau} \mod{\ell}$.

The register $\mathcal{L}_{\tau}^{\xi}$ at a site $\xi$ can take the obvious value $\xi$, but also the value $\xi+\ell$, if the particle has reached the site after having crossed the entire ring once, the value $\xi+2\ell$, if the ring has been crossed twice, etc. Also values $\xi-\ell$, $\xi-2\ell$, $\ldots$ are possible, if the ring is crossed in the opposite direction. Summarizing, $\mathcal{L}_{\tau}^{\xi}$ can take infinitely many distinct values $\xi\pm n\ell$. Consequently, all possible particle paths interefere, with a path difference that is always a multiple of $\ell$. In the proposed model this situation is equivalent to a multiple-source preparation, with infinitely many equally-probable and equally-spaced sources separated by a distance $\ell$ in lattice units. 

In this scenario, the probability density function of the particle momentum is $f(p')=\delta(p'-p)$, that is, an ensemble of particles having the same momentum $p$ is considered. The steady-state probability density of the total momentum $\bp$ is to be calculated. Equation (\ref{eqn:Ep'}) is replaced by
\begin{equation}
  \bp = p- \lim_{N_s\rightarrow\infty} \sum_{i=1}^{N_s} \sum_{j=i+1}^{N_s}  \frac{2}{N_s} \frac{\sin\left(\pi |j-i| \ell q\right)}{\pi |j-i| \ell}.
	\label{eqn:saw}
\end{equation}
The limit at the right-hand side of (\ref{eqn:saw}) reads
\begin{equation}
  \left(\frac{1}{\ell}-\bp+\left\lfloor \bp\frac{\ell}{2} \right\rfloor \right),
\end{equation}
where $\lfloor\cdot\rfloor$ denotes the ``floor" function. Using this result and after manipulation equation (\ref{eqn:saw}) is equivalent to
\begin{equation}
  \bp = \frac{2}{\ell}\left[\frac{p\ell}{2}\right],
\end{equation}
where the operator $[\cdot]$ denotes the rounding function. In other words, for a well-defined particle momentum $p$, the steady-state momentum $\bp$ can only take one of the discrete values $\displaystyle\frac{2n}{\ell}$, that for $n=[\frac{p\ell}{2}]$. This result coincides precisely with QM predictions derived in Appendix~\ref{app:ring}, after transformation to lattice units.

%%
%\begin{equation}
  %f(q) = \lim_{N_s \rightarrow \infty} \frac{1 + \sum_{i=1}^{N_s} \sum_{j=i+1}^{N_s}  \frac{2}{N_s}\cos\left(\pi |j-i| \ell q\right)}{2},
	%\label{eqn:ringlim}
%\end{equation}
%%
%where the fact has been used that the sources are equally probable ($P_i=1/N_s, \forall i$). Taking the limit in (\ref{eqn:ringlim}), the resulting momentum probability density is
%%
%\begin{equation}
  %f(q) = \delta(q-\frac{2n}{\ell}), \quad n=0,\pm 1, \ldots, \pm \ell
	%\label{eqn:ring}
%\end{equation}
%%

%%%%%%%%%%%%%%%%%%%%%%%%%%%%%%%%%%%%%%%%%%%%%%%%
\subsubsection{Particle in a Box} \label{sec:box}
%%%%%%%%%%%%%%%%%%%%%%%%%%%%%%%%%%%%%%%%%%%%%%%%

Although in the ``particle in a box" scenario the particle is not free but submitted to external forces as soon as it hits the box boundaries, this case is easily treated with purely geometric considerations.

The spatial domain is limited to $\xi\in[0,\ell]$, where $\ell X=L$ is the physical width of the box. When the particle hits one of the boundaries, it is ``reflected" due to the infinite potential that is assumed outside the box. In the proposed model, a reflexion (infinite external force) means two things: change of sign of (i) the momentum propensity, of the momentum of all bosons possessed, and of the spatial counter of the particle, and (ii) reset of its time counter to zero. Due to the latter change, the particle can interfere with its previous paths through creation of bosons. In fact, the register $\mathcal{L}_{\tau}^{\xi}$ at a site $\xi$ can take the obvious value $\xi$ (assume for simplicity $\xi_0=0$) if the site is visited before any reflexion, the value $-\ell-(\ell-\xi)$ if it is visited after one reflexion, the value $2\ell+\xi$ if it is visited after two reflexions, etc. Consequently, the path difference is always a multiple of $2\ell$. 

In the proposed model this situation is equivalent to a multiple-source preparation, with infinitely many equally-probable and equally-spaced sources separated by a distance $2\ell$ in lattice units. From considerations that are similar to those presented for the particle in a ring case, it is found that the steady-state probability density of momentum $\bp$ has peaks for the values $\frac{n}{\ell}$, with $n=[p\ell]$. This result coincides precisely with QM predictions derived in Appendix A.6.6, after transformation to lattice units.

%%%%%%%%%%%%%%%%%%%%%%%%%%%%%%%%%%%%%%%%%%%%%%%%%
\subsection{Numerical results}
%%%%%%%%%%%%%%%%%%%%%%%%%%%%%%%%%%%%%%%%%%%%%%%%%

%\subsubsection{Equally-probable sources}
%Instead of attempting deriving an analytical calculation of $\alpha^{(n_J)}$, $V^{(n_J)}$, and thus $\rho(\xi,\tau)$, this section presents numerical result obtained with the pseudocode of Table~\ref{tab:codeint}.

This section is aimed at reproducing numerically the self-interference scenarios analysed in the previous sections. The model described in the previous sections is in principle implemented by the algorithm shown in Table~\ref{tab:codeint}. With respect to the algorithm of Table~\ref{tab:pseudo}, more instructions are now needed to represent the creation of pairs of bosons, their initialization, and the decay of their momentum.

\begin{algorithm}
\caption{Pseudocode used for the simulations of Fig.~\ref{fig:interfTOT}.} \label{tab:codeint}
\begin{algorithmic}[1]
\For{particle 1 to $N_P$}
\State \Comment Initialization: 
\State $\xi \gets$ random value bewteen $-\delta$ and $+\delta$ 
\State $\tau \gets 0 $
\State $p0 \gets$ random value beween -1 and +1 
\State $\lambda \gets 0 $
\ForAll{possible bosons $b$ }
\State $p(b) \gets 0 $
\State $k(b) \gets 0 $
\EndFor 
\For{iteration 1 to $N_T$ }
\State \Comment Particle dynamics: 
\State $\tau \gets \tau+1$ 
\State $p \gets p0-\sum p(b)$ 
\State $\upsilon \gets$ random value +1, 0 or -1 with prob. given by (\ref{eqn:a})
\State $\xi \gets \xi+\upsilon$ 
\State $\lambda \gets \lambda+\upsilon$ 
\State $q \gets \lambda/\tau$ 
\State \Comment Momentum decay for particle bosons, see (\ref{eqn:pdecay}): 
\ForAll {possible bosons $b$ }
\State $k(b) \gets k(b)+1$ 
\State $p(b) \gets p(b)(1-1/2/k(b))$ 
\EndFor
\State \Comment Momentum decay for lattice bosons, see (\ref{eqn:wdecay}): 
\ForAll {sites $s$ }
\ForAll {possible bosons b }
\State $l(s,b) \gets l(s,b)+1$ 
\State $w(s,b) \gets w(s,b)*(1-(w0(s,b)/l(s,b))^2)$ 
\EndFor
\EndFor
\State \Comment Boson creation: 
\State site $s \gets \{\xi,\tau\}$ 
\If {$\mu(s)-\lambda = -\delta$ }
\State boson $b \gets$ ``12" 
\ElsIf { $\mu(s)-\lambda = \delta$ }
\State boson $b \gets$ ``21" 
\Else
\State no boson $b$ 
\EndIf
\State \Comment Boson momentum reinitialization: 
\State $p(b) \gets w(s,b)$ 
\State $k(b) \gets 0$ 
\State $w(s,b) \gets q$ 
\State $w0(s,b) \gets \delta q$ 
\State $l(s,b) \gets 0$ 
\EndFor
\State $\{\lambda,\mu(s)\} \rightleftharpoons \{\mu(s),\lambda\}$ 
\State $\nu(\xi) \gets \nu(\xi)+1$ 
\EndFor
\State $\nu(\xi) \gets \nu(\xi)/N_P$
\end{algorithmic}
\end{algorithm}

\begin{figure}[t!]
  \centering
  \includegraphics[width=\textwidth]{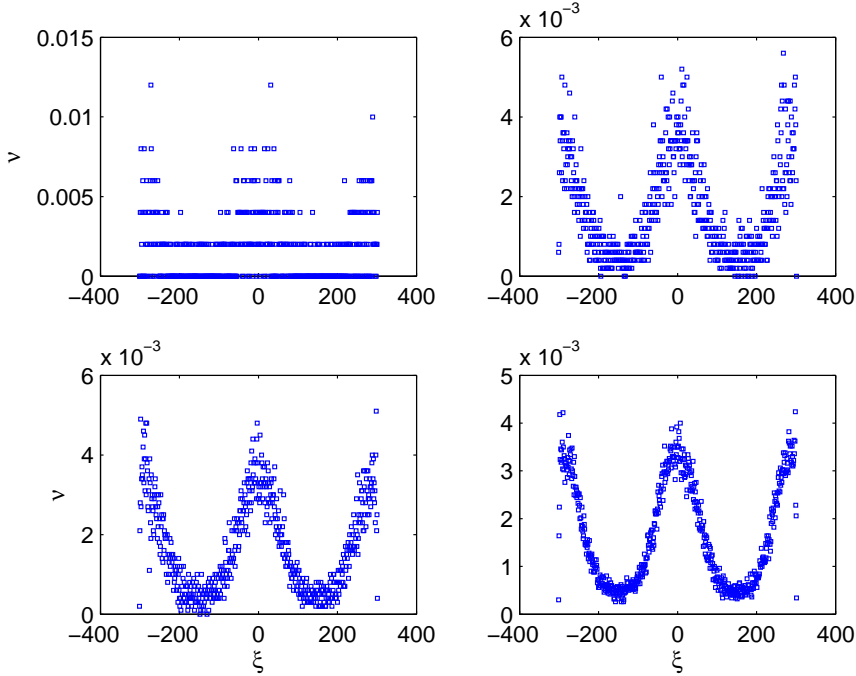}
  \caption{Frequency of arrival of particles emitted at $\xi=\pm 1$ as a function of $\xi$ after $N_T=300$. From top-left to bottom-right, $N_P=$ 500, 5000, 10000, and 50000, respectively.}
  \label{fig:interfTOT}
\end{figure}

Numerical results of Fig.~\ref{fig:interfTOT} have been obtained with a modified version of the algorithm above, aimed at fastening calculations. In fact, the interference pattern starts to build when the frequency of $\lambda\mu$ events approaches the theoretical probabilities $P_{\lambda\mu}$ introduced in Sect.~\ref{sec:apriori}. That requires that a sufficiently large number of lattice sites are visited by a sufficiently large number of particles. This ``lattice training" process requires in turn that a large number, say $N_{P1}$, of particles are emitted. When the number of particle emissions required for the statistical build up of the intereference pattern is considered, it becomes clear that complete simulations would require extensive computing times.

\begin{figure}[t!]
  \centering
  \includegraphics[width=0.5\textwidth]{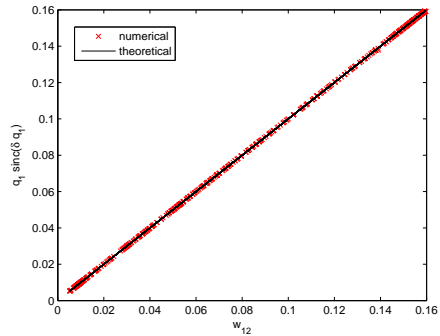}
  \caption{Values of $w_{12}$ in lattice of size $N_T$ after $N_P=100$ particle emissions, vs. respective steady-state values.}
  \label{fig:w12}
\end{figure}

To fasten computations, I have considered the ``lattice training" already completed. In practice, the values $\mu$ found by a particle are not acessed in the site registers but randomly attributed with probabilities that equal the steady-state ones. In the two-slit scenario, $\mu$ equals $q_1$ with probability $P_1$, and $q_2:=(\xi+\delta/2)/\tau$ with probability $P_2$. Similarly, particle bosons are directly initialized with their expected steady-state momenta, see (\ref{eqn:sinc}). In fact, the convergence of the process (\ref{eqn:product}) requires much less iterations than the training of the lattice, and it is completed well before $N_{P1}$ emissions. The latter statement can be observed in Fig.~\ref{fig:w12} that shows pairs of values $w_{12}$ and $q_1{\rm sinc}(2q_1)$ in a lattice after $N_P=100$ and for $N_T=300$ (no gap between two successive emissions).

Figure~\ref{fig:interfTOT} shows the frequency $\nu(\xi)$ calculated for the two-slit scenario ($\delta=2$) after a time $N_T=300$ for different values of $N_{P}$, with the lattice already trained as specified above. As the the number of particles emitted in the ensemble increases, a frequency distribution builds up. For large $N_P$, the frequency clearly tends to the a priori probability $P(\xi,N_T)$ given by (\ref{eqn:p2slit}).

%\subsection{Non equally-probable sources}
%%%%%%%%%%%%%%%%%%%%%%%%%%%%%%%%%%%%%%%%%%%%%%%%%%
%In such a case, the final distribution will be a weighted average of the distributions of the emitted particles that are not subject to the quantum force, and of those who are. For the former, the distribution is given by $P(\xi,\tau)=1/\tau$ (in the large-$\tau$ limit), while for the latter the distribution is given by (\ref{eqn:p2slit}). The weighted average is
%\begin{equation}
  %P(\xi,\tau)= \frac{1-2\sqrt{P_1P_2}}{2\tau} + 2\sqrt{P_1P_2}\cdot\frac{1+\cos 2\pi q}{2\tau} = \frac{1+2\sqrt{P_1P_2}\cos 2\pi q}{2\tau},
%\end{equation}
%which is precisely the same as (\ref{eqn:p2noneq}).
%
%\subsection{Multiple equally-probable sources}
%%%%%%%%%%%%%%%%%%%%%%%%%%%%%%%%%%%%%%%%%%%%%%%%%%
%The extension to any number of sources (slits) is also tedious but straightforward. In definition (\ref{eqn:nu3}) the coefficient 2 is replaced by $n_s$, the number of sources. Since $\Pr(\mu_{\tau}>0)=\frac{n_s-1}{n_s}$, the result (\ref{eqn:Enu3}) becomes $E[\nu_{\tau}^{\xi}]=-(n_s-1)(q_{\tau}-q^{\infty}_{\tau})$. However, the expressions for $q_{\tau}^{\infty}$ are more complex.

\begin{figure}[t!]
  \centering
  \includegraphics[width=\textwidth]{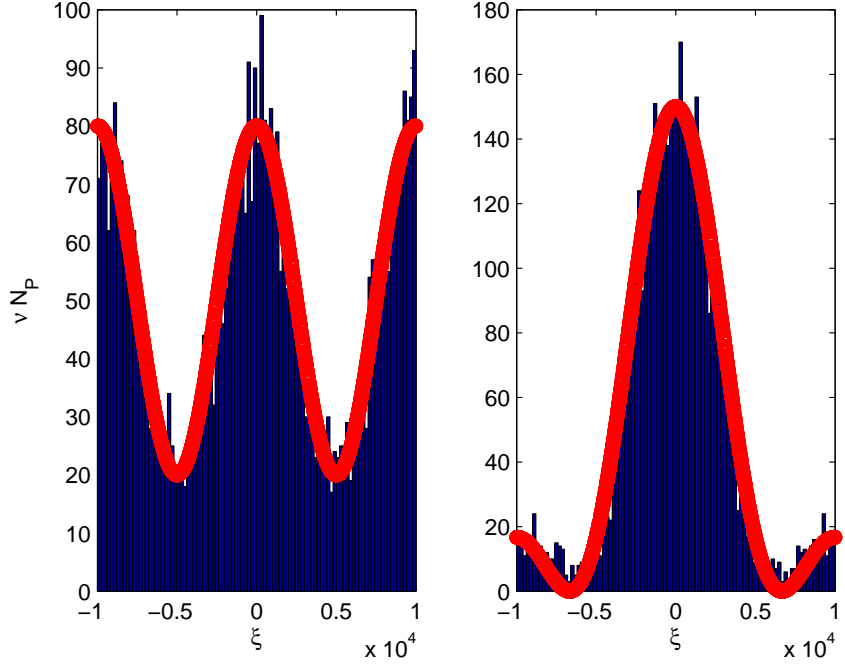}
  \caption{Number of particle arrivals for: two sources at $\xi=\{\pm 1\}$ with respective probability 0.1 and 0.9 (left), three equally-probable sources at $\xi=\{-1,0,1\}$; both scenarios with $N_T=10000$, $N_P=5000$; red curves: theoretical probability densities.}
  \label{fig:interslit}
\end{figure}

Two more scenarios are presented in Fig.~\ref{fig:interslit}. The former is a case of two sources at distance $\delta=2$ with non-equal probabilities $P_1=0.9$, $P_2=0.1$. The second scenario is that of three equally-probable sources located at $\xi=\{-1,0,1\}$. The frequency values have been obtained with the accelerated algorithm discussed above. In the first scenario, two types of bosons are created, both having creation probability $P_1P_2$ and distance $\delta$. In the second scenario, there are two types of bosons with $\delta=2$ and four types of bosons with $\delta=4$, all of them with probability equal to 1/9. In both cases, the figure shows a precise agreement between large $\tau$'s prediction of the model with the theoretical probability densities. 

\section{Discussion and Future Work}
%%%%%%%%%%%%%%%%%%%%%%%%%%%%%%%%%%%%%%%%%%%%%%%%%
%%%%%%%%%%%%%%%%%%%%%%%%%%%%%%%%%%%%%%%%%%%%%%%%%

The proposed approach has been proven capable of describing trajectories of individual particles in an ensemble of similarly-prepared particles in a simple and realistic way. Simple and realistic means that the ontology of the proposed model includes real particles, a real discrete spacetime lattice, both capable of storing and exchanging a few pieces of information, and arithmetic operations. In fact, all quantities modeled are represented either by integers or rational numbers, and operations on them are products and sums.

The predictions of the model have been shown to tend to the predictions of QM in the continuum limit for free particles and, more remarkably, also in the case of quantum interference. In my opinion, the latter evidence makes the model a successful candidate to provide both qualitative and \emph{quantitative} explanation for several quantum phenomena. However, other QM aspects have still to be added to the model. %Three possible extensions are briefly discussed below.

%%%%%%%%%%%%%%%%%%%%%%%%%%%%%%%%%%%%%%%%%%%%%%%%%
%\subsection{Interacting Particles} \label{sec:forces}
%%%%%%%%%%%%%%%%%%%%%%%%%%%%%%%%%%%%%%%%%%%%%%%%%

The results of this paper concern free particles only. However, the proposed model seems naturally capable to integrate also external forces into the picture. Each interaction of the particle with its sourrounding is indeed expected to modify its intrinsic properties. External forces shall be treated as permanent variations of the particle intrinsic momentum propensity $p$, due to new types of \emph{bosons} carrying a specific momentum that amounts to that of the particle. The particle counters shall be reset similarly to the particle-in-a-box scenario. Relativistic Newton's second law shall be considered, to prevent that $p$ becomes larger than unity under the action of persistent forces.
%The rule 
%%
%\begin{equation}
  %q \mapsto p
%\end{equation}
%%
%is assumed and will be tested as a further work.

%A special case of external force is when the particle hits an infinite potential barrier (that is, a site that carries such an information). When that happens, $p$ shall be assumed to instantaneously change its sign, similarly to what happens to the velocity of a classical particle. Moreover, the multiple counters carried by the particle shall be properly reset. 

Extension to two- and three-dimensional spaces seems also natural. A set of three momentum propensities, $p_x$, $p_y$, and $p_z$ shall be introduced, fulfilling the condition that the total energy $e=\frac{1+p_x^2+p_y^2+p_z^2}{2} \leq 1$. This condition implies that $p_x^2+p_y^2+p_z^2 \leq 1$, thus fixing a constraint to the probability densities of the three propensities.

%\end{document}   

\appendix
%%%%%%%%%%%%%%%%%%%%%%%%%%%%%%%%%%%%%%%%%%%%%%%%%
%%%%%%%%%%%%%%%%%%%%%%%%%%%%%%%%%%%%%%%%%%%%%%%%%
\section{Appendices}
%%%%%%%%%%%%%%%%%%%%%%%%%%%%%%%%%%%%%%%%%%%%%%%%%
%%%%%%%%%%%%%%%%%%%%%%%%%%%%%%%%%%%%%%%%%%%%%%%%%

%%%%%%%%%%%%%%%%%%%%%%%%%%%%%%%%%%%%%%%%%%%%%%%%
\subsection{Frequency and matter waves} \label{app:matter}
%%%%%%%%%%%%%%%%%%%%%%%%%%%%%%%%%%%%%%%%%%%%%%%%%
In some interpretations of quantum phenomena, a particle is associated with a matter wave, whose frequency is proportional to its energy via the Planck constant. In the proposed model, the frequency is retrieved as the reciprocal of the \emph{average return time} to any position of the lattice. To see that, define the probability mass function $P(n)$ as the probability that a particle returns at an arbitrary position for the first time after a time $2n$. For example, $P(1)=4ac=b^2/2$, $P(2)=2a^2c^2+2ab^2c=5/8b^4$, etc. The general expression for $P(n)$ is
\begin{equation}
  P(n) = 2b^{2n}\left(\frac{1}{4}+\sum_{k=2}^n \frac{1+4(n-k)}{4^k}\right) = 2b^{2n}\left( \frac{n}{3}-\frac{4}{9}+\frac{13}{9}\left( \frac{1}{4}\right)^n \right)
\end{equation}

\begin{figure}[t!]
  \centering
  \includegraphics[width=\textwidth]{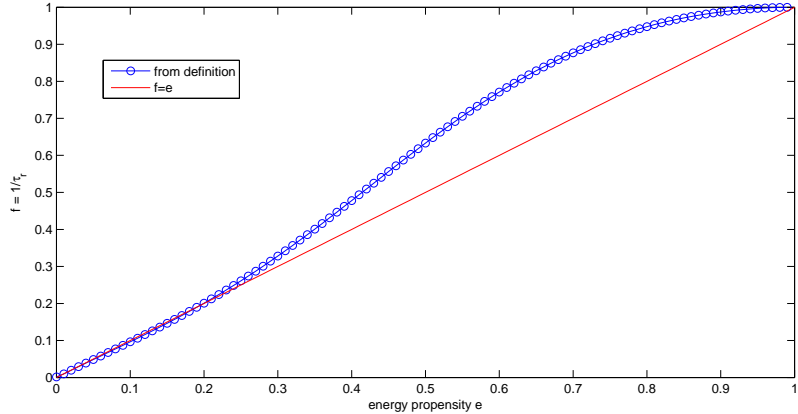}
  \caption{Matter wave frequency as calculated from (\ref{eqn:frequency}) and as $f=e$.}
  \label{fig:matter}
\end{figure}

Now, define the average return time as
\begin{equation}
  \tau_r := E[n] = \frac{\sum_{n=1}^{\infty}nP(n)}{\sum_{n=1}^{\infty}P(n)}
\end{equation}
Using the results
\begin{equation}
  \sum_{k=1}^{\infty} kr^k = \frac{r}{(1-r)^2}, \quad \sum_{k=1}^{\infty} k^2r^k = \frac{r(1+r)}{(1-r)^3},
\end{equation}
one can find that
\begin{equation}
  \sum_{n=1}^{\infty}P(n) = \frac{2}{3}\frac{b^2}{(1-b^2)^2}-\frac{8}{9}\left(\frac{1}{(1-b^2)^2}-1\right)+\frac{26}{9}\left(\frac{1}{(1-b^2/4)^2}-1\right),
	\label{eqn:Pn1}
\end{equation}
\begin{equation}
  \sum_{n=1}^{\infty}nP(n) = \frac{2}{3}\frac{b^2(1+b^2)}{(1-b^2)^3}-\frac{8}{9}\frac{b^2}{(1-b^2)^2}+\frac{26}{9}\frac{b^2/4}{(1-b^2/4)^2},
	\label{eqn:Pn2}
\end{equation}
and consequently $\tau_r$ as a function of $b$. Let me now introduce the energy with the substitution $b=1-e$. After some tedious but straightforward manipulations of (\ref{eqn:Pn1})--(\ref{eqn:Pn2}), find the frequency $f=1/\tau_r$ as
\begin{equation}
  f = \frac{e(2-e)(-1-e)(3-e)(e^4-4e^3+5e^2-2e+1)}{(e^2-2e-1)(5e^4-20e^3+29e^2-18e+6)}
  \label{eqn:frequency}
\end{equation}
that is the relationship sought. It is easy to verify (Fig.~\ref{fig:matter}) that for $e=0$, $f=0$, while for $e=1$, also $f=1$. Moreover, for small values of $e$, the relationship (\ref{eqn:frequency}) is approximated by
\begin{equation}
  f = e,
\end{equation}
which is precisely the de Broglie relation in lattice units.

%%%%%%%%%%%%%%%%%%%%%%%%%%%%%%%%%%%%%%%%%%%%%%%%%
\subsection{Alternative derivations of (\ref{eqn:normal})} \label{app:normal}
%%%%%%%%%%%%%%%%%%%%%%%%%%%%%%%%%%%%%%%%%%%%%%%%%
Consider (\ref{eqn:rhop}) as a binomial distribution $f(k;n,q)$ with $k={\xi}+\tau$, $n=2\tau$, $q=(1+p)/2$. For $n$ large enough an approximation of $\rho$ is a normal distribution with mean $\mu=nq=\tau+\tau p$ and variance $\sigma^2=nq(1-q)=b\tau$. Recalling that $k-\mu={\xi}-p\tau$, obtain (\ref{eqn:normal}). 

Yet a third possible method would start from expressing the recursive equation (\ref{eqn:process}) as 
\begin{equation}
\begin{split}
  \rho_{\tau}(\xi)-\rho_{\tau-1}(\xi) = -p\cdot \frac{\rho_{\tau-1}(\xi+1)-\rho_{\tau-1}(\xi-1)}{2} + \\ +e\cdot \frac{\rho_{\tau-1}(\xi+1)-2\rho_{\tau-1}(\xi)+\rho_{\tau-1}(\xi-1)}{2}.
	\end{split}
\label{eqn:process_diff}
\end{equation}
The latter difference equation has a continuum limit described by the differential equation
\begin{equation}
  \frac{\partial \rho}{\partial \tau} = -p \displaystyle\frac{\partial \rho}{\partial \xi} + e \displaystyle\frac{\partial^2 \rho}{\partial \xi^2}
  \label{eqn:process_pde}
\end{equation}
which is a convective--diffusion equation with $e$ playing the role of the diffusivity and $p$ of the convection velocity. Note, however, that in (\ref{eqn:normal}) the correct result for the diffusivity is $b=1-e$ and not $e$ as it would be predicted by (\ref{eqn:process_pde}).

%%%%%%%%%%%%%%%%%%%%%%%%%%%%%%%%%%%%%%%%%%%%%%%%%
\subsection{Special Relativity} \label{app:specrel}
%%%%%%%%%%%%%%%%%%%%%%%%%%%%%%%%%%%%%%%%%%%%%%%%%

Equations (\ref{eqn:rhop}) or (\ref{eqn:normal}) are invariant with respect to Lorentz transformations. To see that, let me take a particle having momentum propensity $p$ and a reference frame that is moving at velocity $V$ with respect to the fundamental lattice. In this refercne frame, the fundamental lattice dimensions (primed quantities) are deformed as follows:
\begin{equation}
  X' := \frac{X (1-q\beta)\sqrt{1-\beta^2}}{(1-p\beta)^2}, \quad T' := \frac{T (1-q\beta)\sqrt{1-\beta^2}}{(1-p\beta)^2},
\end{equation}
where $q=\frac{\xi}{\tau}$ and $\beta:=V/c$, so that the relationship $X'/T'=X/T=c$ still holds.

Consequently, 
\begin{equation}
  \xi'= \frac{(1-p\beta)^2}{(1-\beta^2)(1-q\beta)}(\xi-\beta\tau), \quad \tau'=\frac{(1-p\beta)^2}{(1-\beta^2)(1-q\beta)}(\tau-\beta\xi)
\end{equation}
so that
\begin{equation}
  x' = \xi' X' = \gamma(x-Vt), \quad t'= \tau' T' = \gamma(t-\frac{Vx}{c^2}),
\end{equation}
with $\gamma=1/\sqrt{1-\beta^2}$, in agreement with Lorentz transformations.

In both reference frames, the probability of having the particle at a site of the fundamental lattice $(\xi,\tau)$ must be the same. In the primed reference frame, (\ref{eqn:normal}) is calculated as
\begin{equation}
  \rho(\xi',\tau') \approx \frac{1}{\sqrt{2\pi b' \tau'}} \exp\left(-\frac{({\xi'}-p'\tau')^2}{2b'\tau'}\right).
 \label{eqn:normalprime}
\end{equation}
Now, let me assume that the momentum propensity in the moving reference frame is
\begin{equation}
  p' = \frac{p-\beta}{1-\beta p}
\end{equation}
in agreement with relativistic velocity addition formula. Consequently, since $b'=(1-p'^2)/2$, one easily verifies that
\begin{equation}
  b' = b\frac{1-\beta^2}{(1-p\beta)^2}.
\end{equation}

%where $q:=\frac{\tilde{\xi}}{\tau}$ is the average speed. The latter equation tends to the relativistic velocity addition formula when $q\approx p$, that is, for large $\tau$'s. 

With these relationships, it is easy verified that
\begin{equation}
  \xi'-p'\tau' = (\xi-p\tau) \frac{1-p\beta}{1-q\beta}
	\label{eqn:xpt}
\end{equation}
and
\begin{equation}
  b'\tau' = b\tau
	\label{eqn:bt}
\end{equation}
Using (\ref{eqn:xpt}) and (\ref{eqn:bt}), it is easily verified that the quantity $\rho(\xi',\tau')$ given by (\ref{eqn:normalprime}) approximates $\rho(\xi,\tau)$ given by (\ref{eqn:normal}) when $q \approx p$, that is, for large $\tau$'s. In reality, all the formulae above are valid for large $\tau$'s, since, for instance, I do not impose that $\xi'$ or $\tau'$ are integers. 

In our laboratory system (primed reference frame moving with an unknown velocity $V$ with respect to the fundamental lattice), we prepare an experiment with a momentum propensity that we label as $p'$ and we observe a probabilty mass function at a point in the spacetime that we label as $x'$, $t'$. Although our assumptions on these values is incorrect, we measure the correct probability mass function using the $\rho$ formula.

If $\beta=p$, that is, in the reference frame of the particle, we obtain the following results: $X'=\gamma X$, $T'=\gamma T$, $\xi'=\xi-p\tau$, $\tau'=\tau-p\xi$, $p'=0$, $b'=1/2$. Thus the particle itself ``sees" a broader lattice ($X'\geq X$, $T' \geq T$). For $\beta=p \rightarrow \pm 1$, the lattice becomes infinitely large.

%%%%%%%%%%%%%%%%%%%%%%%%%%%%%%%%%%%%%%%%%%%%%%%%%
%\subsection{Action} 
%%%%%%%%%%%%%%%%%%%%%%%%%%%%%%%%%%%%%%%%%%%%%%%%%

%%%%%%%%%%%%%%%%%%%%%%%%%%%%%%%%%%%%%%%%%%%%%%%%%
\subsection{Derivation of (\ref{eqn:PrS})} \label{app:phis}
%%%%%%%%%%%%%%%%%%%%%%%%%%%%%%%%%%%%%%%%%%%%%%%%%
The number of paths leading to a certain site is given by 
\begin{equation}
  N_p = \sum_{n_a} \sum_{n_b} \sum_{n_c} {\tau \choose n_a} \cdot {\tau-n_a \choose n_b} \cdot {\tau-n_a-n_b \choose n_c}
  \label{eqn:npath}
\end{equation}
where $n_a$, $n_b$, and $n_c$ are the number of moves with $\upsilon=1,0,-1$, respectively. Clearly, $n_a+n_b+n_c=\tau$ and, in order to reach exactly the site in question, $n_a-n_c=\xi$. With these constraints, (\ref{eqn:npath}) becomes
\begin{equation}
  N_p = \sum_{n_c} {\tau \choose n_c+\xi} \cdot {\tau-n_c-{\xi} \choose \tau-2n_c-{\xi}} := \sum_{n_c} N_p(n_c)
  \label{eqn:npath1}
\end{equation}
The range of $n_c$ derives from the fact that all the binomial arguments are positive integers. Therefore, $n_c+{\xi} \geq 0$, $\tau-n_c-{\xi} \geq 0$, $\tau-2n_c-{\xi} \geq 0$. Consequently, $\max(0,-{\xi}) \leq n_c \leq \left\lfloor \frac{\tau-{\xi}}{2} \right\rfloor$. 

The probability of each path is given by $a^{n_a}b^{n_b}c^{n_c}$. Since $b^2=4ac$, this probabilty is also calculated as $2^{n_b}a^{n_a+n_b/2}c^{n_c+n_b/2}$. From (\ref{eqn:rhop}), the probability of \emph{all} $N_p$ paths is
\begin{equation}
  \displaystyle\frac{{2\tau \choose \tau+{\xi}}}{2^{2\tau}}(1+p)^{\tau+{\xi}}(1-p)^{\tau-{\xi}} = {2\tau \choose \tau+{\xi}} a^{\frac{\tau+{\xi}}{2}} c^{\frac{\tau-{\xi}}{2}} = {2\tau \choose \tau+{\xi}} a^{n_a+n_b/2}c^{n_c+n_b/2}.
\end{equation}

For each value of $n_c$ there will be a number $N_p(n_c)$ of paths with the same cumulated energy. The value taken by $\mathcal{S}_{\tau}^{\xi}$ equals the sum $n_a+n_c=2n_c+{\xi}$ since only these moves contribute to the cumulated energy. The relative probability of these $N_p(n_c)$ paths over the totality of $N_p$ paths is thus (\ref{eqn:PrS}).

For the sake of completeness, one might define a particle stochastic variable as well,
\begin{equation}
  \mathcal{S}_{n}:=\sum_{n'=n_0}^{n}|\mathcal{V}_{n'}|=\sum_{n'=n_0}^{n}\mathcal{V}_{n'}^2,% =\sum_{n'=n_0}^{n}\mathcal{E}_{n'}.
	\label{eqn:Sstoc}
\end{equation}
whose pmf follows the binomial distribution with $\tau$ trials and $e=a+c$ probability of success, i.e.
\begin{equation}
  \phi_{\tau}(\sigma) := \Pr(\mathcal{S}_{\tau}=\sigma) = {\tau \choose \sigma} e^{\sigma} b^{\tau-\sigma}.
\end{equation}
with support $\sigma=\{0,\ldots,\tau\}$ ($e$ is the energy propensity, not the Neper number).

%%%%%%%%%%%%%%%%%%%%%%%%%%%%%%%%%%%%%%%%%%%%%%%%
\subsection{Derivation of (\ref{eqn:P})} \label{app:P}
%%%%%%%%%%%%%%%%%%%%%%%%%%%%%%%%%%%%%%%%%%%%%%%%%

From the definition (\ref{eqn:intp}),
\begin{equation}
  P_{\tau}^{\xi} = \frac{\left(
\begin{array}{c}
	2\tau \\ \tau+\xi
\end{array}
 \right)}{2^{2\tau+1}}2^{2\tau+1}B(\tau+\xi+1,\tau-\xi+1) 
  \label{eqn:prob}
\end{equation}
where $B(\cdot)$ is here the Beta function,
\begin{equation}
  B(v,w) := \int_0^1 z^{v-1}\cdot (1-z)^{w-1} dz.
\end{equation}
From the properties of such function, it follows that
\begin{equation}
  P_{\tau}^{\xi} =  \left(
\begin{array}{c}
	2\tau \\ \tau+\xi
\end{array}
 \right)\frac{(\tau+\xi)!(\tau-\xi)!}{(2\tau+1)!},
\end{equation}
that is, (\ref{eqn:P}).

%%%%%%%%%%%%%%%%%%%%%%%%%%%%%%%%%%%%%%%%%%%%%%%%
\subsection{Probability densities for the scenarios considered} \label{app:single}
%%%%%%%%%%%%%%%%%%%%%%%%%%%%%%%%%%%%%%%%%%%%%%%%%

\subsubsection{Single source}
%%%%%%%%%%%%%%%%%%%%%%%%%%%%%%%%%%%%%%%%%%%%%%%%%

The wavefunction for a free particle is
\begin{equation}
  \Psi(x,t) = \frac{1}{\sqrt{2\pi}} \int e^{i(kx-\omega t)}\varphi(k,0)dk,
  \label{eqn:psi}
\end{equation}
where the wavenumber $k$ is related to the momentum of the particle and
\begin{equation}
  \varphi(k,0) = \frac{1}{\sqrt{2\pi}} \int \Psi(x',0)e^{-ikx'}dx'.
  \label{eqn:phi}
\end{equation}
For a single perfectly localized source at $x=0$, (\ref{eqn:phi}) reads
\begin{equation}
  \varphi(k,0) = \frac{X}{\sqrt{2\pi}}
  \label{eqn:phi0}
\end{equation}
and consequently (\ref{eqn:psi}) is rewritten as
\begin{equation}
  \Psi(x,t) = \frac{X}{\sqrt{2\pi}} \int \exp\left[i(kx-\frac{\hbar k^2}{2m} t)\right]dk
  \label{eqn:psi1}
\end{equation}
that, integrated, yields
\begin{equation}
  \Psi(x,t) = \frac{X}{\sqrt{2\pi}}\sqrt{\frac{m}{i\hbar t}}\exp\left[i\frac{mx^2}{2\hbar t}\right].
  \label{eqn:psi2}
\end{equation}
The probability density is easily calculated as
\begin{equation}
  |\Psi(x,t)|^2 = \frac{mX^2}{2\pi\hbar t} = \frac{mX^2}{ht}
\end{equation}

\subsubsection{Equally-probable sources}
%%%%%%%%%%%%%%%%%%%%%%%%%%%%%%%%%%%%%%%%%%%%%%%%%
The solution of the Schr\"{o}dinger equation for the case where the two sources are equally probable is based on the linear superimposition of the two waveforms, 
\begin{equation}
  \Psi(x,t) = \frac{\Psi_1(x,t)+\Psi_2(x,t)}{\sqrt{2}},
	\label{eqn:psi2slit}
\end{equation}
where both $\Psi_1(x,t)$ and $\Psi_2(x,t)$ are obtained from (\ref{eqn:psi2}) by replacing  the term $x^2$ (that was valid for a source at $x=0$) with a term $(x-\frac{\delta}{2} X)^2$ and $(x+\frac{\delta}{2} X)^2$, respectively. Thus,
\begin{equation}
  \Psi(x,t) = \frac{X}{2\sqrt{\pi}}\sqrt{\frac{m}{i\hbar t}}\left\{\exp\left[i\frac{m(x-\frac{\delta}{2} X)^2}{2\hbar t}\right]+\exp\left[i\frac{m(x+\frac{\delta}{2} X)^2}{2\hbar t}\right]\right\}.
  \label{eqn:2psi}
\end{equation}
The probability density is given by
\begin{equation}
  |\Psi(x,t)|^2 = \frac{2mX^2}{h t}\cos^2\left(\frac{S_1-S_2}{2}\right),
\end{equation}
where $S_1$ and $S_2$ are the two independent action values, that is, the phases of the two exponentials in (\ref{eqn:2psi}). As $S_1-S_2 = \displaystyle\frac{2\pi m\delta X x}{ht}$, obtain
\begin{equation}
  |\Psi(x,t)|^2 = \frac{mX^2}{ht} \left(1+\cos \left( 2\pi \displaystyle\frac{m\delta X x}{ht}\right) \right)
  \label{eqn:p2slit}
\end{equation}
and thus an interference term arises with respect to (\ref{eqn:f0}), due to the presence of two possible sources. The interference is related to the phase difference between the two waveforms.

\subsubsection{Non equally-probable sources} \label{app:2slit}
%%%%%%%%%%%%%%%%%%%%%%%%%%%%%%%%%%%%%%%%%%%%%%%%%
In the case the probability of the two sources is different, say, $P_1 \neq P_2$, equation (\ref{eqn:psi2slit}) is replaced by $\Psi(x,t) = \sqrt{P_1}\Psi_1(x,t)+\sqrt{P_2}\Psi_2(x,t)$. Consequently, (\ref{eqn:p2slit}) is replaced by
\begin{equation}
  |\Psi(x,t)|^2 = \frac{mX^2}{ht} \left(1+ 2\sqrt{P_1 P_2}\cos \left( 2\pi \displaystyle\frac{m\delta X x}{ht}\right) \right)
  \label{eqn:p2slitnoneq}
\end{equation}
If $P_1=P_2=1/2$, (\ref{eqn:p2slit}) is retrieved.

\subsubsection{Multiple sources} \label{app:multi}
%%%%%%%%%%%%%%%%%%%%%%%%%%%%%%%%%%%%%%%%%%%%%%%%%
In the case of $N_s$ sources, (\ref{eqn:p2slit}) is generalized as follows.
\begin{equation}
  |\Psi(x,t)|^2 = \frac{mX^2}{ht} \left(1+ \sum_{i=1}^{N_s} \sum_{j=i+1}^{N_s} 2\sqrt{P_iP_j}\cos \left( 2\pi \displaystyle\frac{m |x_i-x_j| x}{ht}\right) \right),
  \label{eqn:pNslit}
\end{equation}
%\begin{equation}
  %P(\xi,\tau) = \frac{1+\sum_{i,j,i\neq j} 2\sqrt{P_iP_j}\cos\left(\pi |x_i-x_j| \displaystyle\frac{\xi}{\tau}\right)}{2\tau},
	%\label{eqn:p2mult}
%\end{equation}
%
where $x_i$ is the locations of the $i$-th source and $\sum_{i=1}^{N_s}P_i=1$. If the sources are equally probable and equally spaced by a distance $\delta X$, (\ref{eqn:p2slit}) is particularized as
\begin{equation}
  |\Psi(x,t)|^2 = \frac{mX^2}{ht} \left(1+ \sum_{j=1}^{N_s-1} 2 \frac{N_s-j}{N_s} \cos \left( 2\pi \displaystyle\frac{m j\delta X x}{ht}\right) \right).
  \label{eqn:pNsliteq}
\end{equation}

\subsubsection{Particle in a Ring} \label{app:ring}
%%%%%%%%%%%%%%%%%%%%%%%%%%%%%%%%%%%%%%%%%%%%%%%%%
The ``particle in a ring" scenario can be described by assuming a single plane wave of a well-defined wavenumber and momentum.
%interference between two plane waves, with opposite wavenumbers, representing the two possible directions of motion along the ring. 
Equation (\ref{eqn:psi}) is still valid with $\varphi(k,0)=\delta(k)$,
%+\delta(-k)$, 
where $\delta$ is here the Dirac delta function (assume positive direction of momentum). As the term $e^{i\omega t}$ does not affect the amplitude of the wavefunction, it can be neglected to write
\begin{equation}
 \Psi(x) = \frac{1}{\sqrt{2\pi}} e^{ikx}.%+e^{-ikx})
\label{eqn:psiring}
\end{equation}
However, periodicity imposes that $\Psi(x)=\Psi(x\pm 2\pi R$), where $R$ is the ring radius, or $e^{ikx}=e^{ik(x+2\pi R)}$, which implies that $e^{ik2\pi R}=1$. This condition is satisfied for $2\pi kR=2\pi n$, where $n\in \mathbb{Z}$, or for 
\begin{equation}
 k_n = \frac{n}{R},
 \label{eqn:kn}
\end{equation}
in agreement with the somehow standard derivation that considers angular momentum and polar coordinates. In terms of momentum $p=\hbar k$ (here the dimensional momentum, not the momentum propensity as in the paper body), (\ref{eqn:kn}) reads
\begin{equation}
 p_n = \frac{nh}{L},
 \label{eqn:pn}
\end{equation}
where $L=2\pi R$ is the ring circumference.

%In lattice units ($k=\frac{2\pi}{h}p\frac{mX}{T}$), condition (\ref{eqn:kn}) reads
%%
%\begin{equation}
 %p_n = \frac{n}{\pi r},
%\end{equation}
%%
%where $r=R/X$.

\subsubsection{Particle in a Box} \label{app:box}
%%%%%%%%%%%%%%%%%%%%%%%%%%%%%%%%%%%%%%%%%%%%%%%%%
The wavefunction has the same form (\ref{eqn:psiring}) of the ``particle in a ring", however, with opposite amplitudes for the forward and backward waves,
\begin{equation}
 \Psi(x) = \frac{1}{\sqrt{2\pi}}(e^{ikx}-e^{-ikx}),
\label{eqn:psiring}
\end{equation}
satisfying the boundary condition $\Psi(0)=0$. The second boundary condition $\Psi(L)=0$ imposes that $e^{ikL}=e^{-ikL}$ or $e^{2ikL}=1$. The latter condition is satisfied for $2kL=2\pi n$, where $n\in \mathbb{Z}$, or for
\begin{equation}
 k_n = \frac{n \pi}{L}.
 \label{eqn:kn_box}
\end{equation}
In terms of dimensional momentum, $p_n=\displaystyle\frac{nh}{2L}$ holds.

\subsubsection{Step Potential} \label{app:step}
%%%%%%%%%%%%%%%%%%%%%%%%%%%%%%%%%%%%%%%%%%%%%%%%%

	\begin{table}
		\centering
			\begin{tabular}{lllll}
				Var. & Support & Exp. value & Variance & pmf \\ \hline
				$\mathcal{V}_n$ & $\upsilon\in\{0,\pm 1\}$ & $p$ & $\frac{1-p^2}{2}$ & $\{a,b,c\}$ \\
				$\mathcal{X}_n$ & $\xi\in\{\xi_0-\tau,\ldots,\xi_0+\tau\}$ & $\xi_0+p\tau$ & $\frac{1-p^2}{2}\tau$ & $\rho_n(\xi)$ \\
				$\mathcal{L}_n$ & $\lambda\in\{-\tau,\ldots,\tau\}$ & $p\tau$ & $\frac{1-p^2}{2}\tau$ & $\rho_n(\xi_0+\lambda)$ \\
				%$\mathcal{Q}_n$ & $q\in\{-1,-1+1/\tau,\ldots,1\}$ & $p$ & $\frac{1-p^2}{2\tau}$ & $\rho_n(\xi_0+q\tau)$ \\
				$\mathcal{X}_{\tau}^{\xi}$ & $\xi$ & $\xi$ & 0 & Deter. \\
				$\mathcal{O}_{\tau}^{\xi}$ & $\{0,1\}$ & $\rho_{\tau}(\xi)$ & $\rho_{\tau}(\xi)\left(1-\rho_{\tau}(\xi)\right)$ & $\{\rho_{\tau}(\xi),1-\rho_{\tau}(\xi)\}$ \\
				$\mathcal{S}_n$ & $\sigma\in\{0,\ldots,\tau\}$ & $\frac{1+p}{2}\tau$ & $\frac{1+p}{2}\left(1-\frac{1+p}{2}\right)\tau$ & $\phi_{\tau}(\sigma)$ \\
				$\mathcal{S}_{\tau}^{\xi}$ & $\sigma\in\{|\xi|,|\xi|+2,\ldots,|\xi|+2\left\lfloor \frac{\tau-|\xi|}{2} \right\rfloor\}$ & $\frac{\xi^2+\tau^2-\tau}{2\tau-1}$ & (\ref{eqn:VarL}) & $\phi_{\tau}^{\xi}$ \\ \hline
			\end{tabular}
	\end{table}
	
\subsection{Derivation of (\ref{eqn:sinc})} \label{app:W}
%%%%%%%%%%%%%%%%%%%%%%%%%%%%%%%%%%%%%%%%%%%%%%%%%

I will consider 12-bosons and omit all boson (``12") and site ($\xi,\tau$) indices for the sake of simplicity. 
The boson momentum at a site $\{\xi,\tau\}$ is a stochastic variable depending on its lifetime, another stochastic variable, according to the update rule
\begin{equation}
  \mathcal{W} \leftarrow \left\{\begin{array}{ll}
	w^{(0)}, & \quad \Pr = r' \\ \mathcal{W} \cdot \left(1-\left(\displaystyle\frac{\delta w^{(0)}}{\ell}\right)^2\right) & \quad \Pr=1-r' \end{array} \right.
\end{equation}
\begin{equation}
  \ell \leftarrow \left\{\begin{array}{ll}
	0, & \quad \Pr = r' \\ \ell+1 & \quad \Pr=1-r' \end{array} \right.
\end{equation}
where the probability $r'$ that a new site boson is created is the product of three terms: the frequency with which particles in the ensemble are emitted, $\rho_{\xi}^{\tau}$ (the probability that the site is visited), and $P_{12}$ (the probability of the event 12).

The quantity $w^{(0)}$ is defined by rule (\ref{eqn:wboson0}). Clearly, it is constant after that the site is visited for the first time and 
\begin{equation}
  w^{(0)} = q_{1}
\end{equation}
holds, with $q_1:=\left(\frac{\xi-\delta/2}{\tau} \right)$. Thus the quantities $w^{(\ell)}$ are also constant and can be calculated by repeated application of rule (\ref{eqn:wdecay}),
\begin{equation}
  w^{(\ell)} = q_{1} \prod_{j=1}^{\ell} \left(1-\frac{(\delta q_1)^2}{j^2} \right).
\label{eqn:product}
\end{equation}

The probability that the site carries a boson with lifetime $\ell$ and momentum $w^{(\ell)}$ equals the product $r'(1-r')^{\ell}$. Consequently, the expected value of $\mathcal{W}$ is
\begin{equation}
  E[\mathcal{W}] = r'\sum_{\ell=0}^{\tau}(1-r')^{\ell} q_{1} \prod_{j=1}^{\ell} \left(1-\frac{(\delta q_1)^2}{j^2} \right).%q_1 {\rm Sinc}(\delta q_1)
	\label{eqn:EW}
\end{equation}
Since the values of $r'$ are generally very small, (\ref{eqn:EW}) is rewritten for large $\tau$'s as
\begin{equation}
  E[\mathcal{W}] = q_{1} \prod_{j=1}^{\infty} \left(1-\frac{(\delta q_1)^2}{j^2} \right) = q_1 {\rm Sinc}(\delta q_1).
	\label{eqn:EWstat}
\end{equation}
However, for large $\tau$'s $q_1 \approx \frac{\xi}{\tau}$, which is true also for $q_2:=\left(\frac{\xi+\delta/2}{\tau} \right)$. Therefore, for both bosons ``12" and ``21", 
\begin{equation}
  E[\mathcal{W}] = \frac{\sin(\pi\delta\frac{\xi}{\tau})}{\pi\delta}
	\label{eqn:sin}
\end{equation}
holds.

\subsection{Derivation of (\ref{eqn:Ep12})} \label{app:Pt}
%%%%%%%%%%%%%%%%%%%%%%%%%%%%%%%%%%%%%%%%%%%%%%%%%

The probability that the particle carries a boson ``12" equals the probability that it carries a new boson, \emph{or} a boson that is one iteration-old, \emph{or} a boson that is two iterations-old, etc., all these events being mutually exclusive. Such probability is therefore calculated as
\begin{equation}
  \Pr[\mathrm{boson} ``12"=\mathrm{True}]=P_{12}+P_{12}(1-P_{12})+P_{12}(1-P_{12})^2+\ldots = P_{12} \sum_{k=0}^{\tau-1} (1-P_{12})^k
	\label{eqn:prboson}
\end{equation}
However, we consider large $\tau$'s probabilities to retrieve $P(\xi,\tau)$, so let $\tau$ in (\ref{eqn:prboson}) tend to infinity. 

Boson lifetime affects the momentum exchange, as per (\ref{eqn:pdecay}). Thus the mean value of the boson momentum is calculated as
\begin{equation}
  E[\mathcal{P}_{12}] = P_{12} \sum_{k=0}^{\infty} (1-P_{12})^k p_{12}^{(k)}
	\label{eqn:p12}
\end{equation}
From the definition (\ref{eqn:pdecay}),
\begin{equation}
  \frac{p_{12}^{(k)}}{p_{12}^{(0)}} = \prod_{l=1}^{k} \frac{2l-1}{2l} = \frac{1\cdot 3\cdot 5\cdot 7\cdot \ldots}{2\cdot 4\cdot 6\cdot \ldots} = \frac{(2k)!}{(k!)^2 4^k}.
\end{equation}
After some manipulations, the latter formula turns out to be formally equivalent to
\begin{equation}
  \frac{p_{12}^{(k)}}{p_{12}^{(0)}} = (-1)^k {-1/2 \choose k}
\end{equation}
as it can be verified by inspection. Therefore, (\ref{eqn:p12}) is rewritten as
\begin{equation}
  \frac{E[\mathcal{P}_{12}]}{p_{12}^{(0)}} = P_{12} \sum_{k=0}^{\infty} (1-P_{12})^k (-1)^k {-1/2 \choose k} = P_{12} (1-(1-P_{12}))^{-1/2} = \sqrt{P_{12}}
	\label{eqn:sqrt}
\end{equation}
after having recognized in the right-hand side the binomial series 
\begin{equation}
  (1-x)^{\alpha} = \sum_{k=0}^{\infty} {\alpha \choose k} (-x)^k
\end{equation}
multiplied by $P_{12}$.

The quantity $p_{12}^{(0)}$ is a stochastic variable on its own, since it takes a different value (that of the resident boson) for each site visited when a new pair of bosons is created. From (\ref{eqn:pboson0}) and (\ref{eqn:sinc}) this stochastic variable tends to be distributed as
\begin{equation}
  p_{12}^{(0)} \approx \frac{\sin(\pi \delta \frac{\mathcal{X}_n}{\tau})}{\pi\delta }.
\end{equation}
%
%Consider now the variable
%%
%\begin{equation}
  %q:= \lim_{\tau\rightarrow\infty} E[\mathcal{Q}_{n}].
%\end{equation}
%%
%Using (\ref{eqn:p}), (\ref{eqn:sde}), and replacing $p$ with $p_t$, it turns out (with a certain abuse of notation) that
%%
%\begin{equation}
  %dq = \frac{p_t-q}{\tau}d\tau.
%\end{equation}
%%
%In particular, $q \rightarrow p_t$ for large $\tau$'s. Replacing this value into (\ref{eqn:Ep1}), obtain

With this result, equation (\ref{eqn:sqrt}) is rewritten as
\begin{equation}
  E[\mathcal{P}_{12}] = \sqrt{P_1P_2} \cdot \frac{\sin(\pi \delta \frac{\mathcal{X}_n}{\tau})}{\pi\delta }.
\end{equation}

\end{document}